\documentclass[conference]{IEEEtran}
\IEEEoverridecommandlockouts

\usepackage[textsize=tiny]{todonotes}
\usepackage[utf8]{inputenc}
\usepackage[T1]{fontenc}
\usepackage{microtype}
\usepackage{color}
\usepackage{booktabs}
\usepackage{lipsum}
\usepackage{amsthm}
\newtheorem{theorem}{Theorem}

\theoremstyle{definition}

\usepackage{graphicx}
\usepackage{balance}  

\usepackage[tight,footnotesize]{subfigure}
\usepackage{multirow}
\usepackage{xspace}
\usepackage[ruled,vlined,linesnumbered]{algorithm2e}
\usepackage{color}
\usepackage{listings}
\usepackage{amsmath}
\usepackage{mathtools}
\usepackage{subcaption}
\usepackage{float}

\usepackage[pdfa]{hyperref}
\usepackage{amsmath}
\usepackage{enumitem}
\usepackage{subscript}
\usepackage[skip=0pt]{caption}
\usepackage{ifthen}
\usepackage{tabularx,ragged2e}
\usepackage{graphicx} 
\usepackage{colortbl}
\usepackage{tabulary}
\usepackage{xcolor}
\colorlet{shadecolor}{gray!10}
\usepackage{framed}
\usepackage{enumitem}
\usepackage{makecell}
\usepackage{longtable}
\usepackage{tabu}
\usepackage{tabularx}
\usepackage{tabularray}
\usepackage{caption}
\usepackage{rotating}
\usepackage{adjustbox}
\usepackage{amsmath,pict2e,picture}
\usepackage{amsfonts}
\usepackage{comment}
\usepackage{array}
\usepackage{cuted}
\usepackage[listings,skins,breakable]{tcolorbox}
\newcommand{\probP}{\text{I\kern-0.15em P}}

\newcommand*\circled[1]{\tikz[baseline=(char.base)]{
            \node[shape=circle,fill,inner sep=2pt] (char) {\textcolor{white}{#1}};}}

\newcolumntype{?}{!{\vrule width 1pt}}
\newcommand{\sln}{2}
 
\newcommand{\cm}{\checkmark}

\newcommand{\CactusDB}{\textsc{CactusDB}\xspace}

\usepackage{cite}
\usepackage{amsmath,amssymb,amsfonts}
\usepackage{algorithmic}
\usepackage{graphicx}
\usepackage{textcomp}
\usepackage{xcolor}
\definecolor{MyDarkGray}{RGB}{120,120,120}
\def\BibTeX{{\rm B\kern-.05em{\sc i\kern-.025em b}\kern-.08em
    T\kern-.1667em\lower.7ex\hbox{E}\kern-.125emX}}

\newcommand{\eat}[1]{}

\newcommand{\showComments}{yes}
\newcommand{\submit}{yes} 

\newcommand{\note}[2]{
  \ifthenelse{\equal{\submit}{yes}}{}{%
    \ifthenelse{\equal{\showComments}{yes}}{\textcolor{#1}{#2}}{}
  }
}

\newcommand{\lulu}[1]{\textcolor{teal}{(Lulu:#1)}}

\usepackage{relsize,amsmath} 

\lstdefinestyle{sqlstyle}{
    language=SQL,
    basicstyle=\ttfamily\small,
    keywordstyle=\color{blue},
    commentstyle=\color{gray},
    stringstyle=\color{red},
    numbers=left,
    numberstyle=\tiny\color{gray},
    stepnumber=1,
    numbersep=5pt,
    showstringspaces=false,
    breaklines=true,
    frame=single,
    captionpos=b
}

\lstdefinestyle{pythonstyle}{
    language=Python,
    basicstyle=\ttfamily\small,
    keywordstyle=\color{blue},
    commentstyle=\color{gray},
    stringstyle=\color{red},
    numbers=left,
    numberstyle=\tiny\color{gray},
    stepnumber=1,
    numbersep=5pt,
    showstringspaces=false,
    breaklines=true,
    frame=single,
    captionpos=b
}
\newcommand{\specialcell}[2][c]{%
\begin{tabular}[#1]{@{}c@{}}#2\end{tabular}}

\begin{document}

\title{\LARGE \CactusDB: Unlock Co-Optimization Opportunities \\for SQL and AI/ML Inferences \\
\large (Accepted to ICDE 2026 as a full research paper)
}
\vspace{-25pt}
\author{
Lixi Zhou$^{a}$, Kanchan Chowdhury$^{a}$, Lulu Xie$^{a}$, Jaykumar Tandel$^{a}$, \\
Hong Guan$^{a}$, Zhiwei Fan$^{b}$, Xinwei Fu$^{c}$, Jia Zou$^{a}$ \\ \vspace{-10pt}
\\[-4pt]
{\small $^{a}$ Arizona State University, $^{b}$ Meta, $^{c}$ Amazon}
}

\twocolumn
\maketitle

\begin{abstract}
  There is a growing demand for supporting \textit{inference queries} that combine \textcolor{black}{Structured Query Language (SQL)} and \textcolor{black}{Artificial Intelligence / Machine Learning (AI/ML)} model inferences in database systems, to avoid data denormalization and transfer, facilitate management, and alleviate privacy concerns. Co-optimization techniques for executing inference queries in database systems without accuracy loss fall into four categories: (O1) Relational algebra optimization treating AI/ML models as black-box user-defined functions (UDFs); (O2) Factorized AI/ML inferences; (O3) Tensor-relational transformation; and (O4) General cross-optimization techniques. However, we found none of the existing database systems support all these techniques simultaneously, resulting in suboptimal performance. 
In this work, we identify two key challenges to address the above problem: (1) the difficulty of unifying all co-optimization techniques that involve disparate data and computation abstractions in one system; and (2) the lack of an optimizer that can effectively explore the exponential search space.
To address these challenges, we present \CactusDB, a novel system built atop Velox - a high-performance, UDF-centric database engine, open-sourced by Meta. \CactusDB features a three-level \textcolor{black}{Intermediate Representations (IR)} that supports relational operators, expression operators, and ML functions to enable flexible optimization of arbitrary sub-computations. Additionally, we propose a novel Monte-Carlo Tree Search (MCTS)-based optimizer with query embedding, co-designed with our unique three-level IR, enabling shared and reusable optimization knowledge across different queries.
Evaluation of $12$ representative inference workloads and $2,000$ randomly generated inference queries on well-known datasets, such as MovieLens and TPCx-AI, shows that \CactusDB achieves up to $441\times$ speedup compared to alternative systems.
\end{abstract}

\begin{IEEEkeywords}
database, machine learning, query optimization
\end{IEEEkeywords}

\lstnewenvironment{DSL}
  {\lstset{
        aboveskip=5pt,
        belowskip=5pt,
        mathescape=true,
        basicstyle=\ttfamily\small,
        morekeywords={Set, Vector, Map, HashMap, bool, select,
  multiselect, aggregate, join, partition, member, method, construct, true, false, if, else, CREATE TABLE, PARTITION BY,
  self, literal, vector, return, for push_back, function, enum, sort, string, double},
        literate={~} {$\sim$}{1},
        showstringspaces=false}\vspace{0pt}%
   \noindent\minipage{0.47\textwidth}}
   {\endminipage\vspace{0pt}}

\lstnewenvironment{SQL}
  {\lstset{
        aboveskip=5pt,
        belowskip=5pt,
        escapechar=!,
        mathescape=true,
        language=SQL,
        basicstyle=\linespread{0.94}\ttfamily\small,
        morekeywords={JOIN, FROM, WHERE, SELECT, pgml, project_name, task, relation_name, y_column_name, algorithm, hyperparams},
        deletekeywords={VALUE, PRIOR},
        showstringspaces=false}
        \vspace{0pt}%
        \noindent\minipage{0.47\textwidth}}
  {\endminipage\vspace{0pt}}

\newcommand{\littlesection}[1]{\vspace{5pt}\noindent\textbf{#1}}
\vspace{-5pt}
\section{Introduction}
\label{sec:intro}
It is important to nest \textcolor{black}{Artificial Intelligence / Machine Learning (AI/ML)} inferences and \textcolor{black}{Structured Query Language (SQL)} queries to drive high-value insights from relational data, e.g., to generate personalized product recommendations~\cite{ruan2023dynamic, patel2024lotus, liu2024declarative} based on \textcolor{black}{preprocessed customer profiles and online shopping records through \texttt{join}, \texttt{filter}, and \texttt{project} nested with AI/ML model inferences}, to obtain statistics about credit card transactions that are predicted as fraud~\cite{fraud-ibm}, \textcolor{black}{through \texttt{aggregate},\texttt{join}, and \texttt{filter}, nested with inferences}. A broad class of works on using AI to enhance database systems further strengthens such demands~\cite{kraska2018case, sharma2023automatic, zhou2024deepmapping, guan2024idnet}.
Therefore, a recent trend of supporting SQL queries nested with AI/ML inference functions within database systems has emerged~\cite{gaussml, kakkar2023eva, postgresml, zhou2024serving, lin2023smartlite, shahrokhi2024pytond, shaikhha2021intermediate, chowdhury2025inferf, guan2025declarative, chowdhury2025exboost, yuan2020tensor}. This approach eliminates the need to transfer data from databases to ML systems. It not only reduces latency in workloads where data transfer becomes a bottleneck~\cite{guan2023comparison} but also alleviates privacy concerns~\cite{annas2003hipaa} and operational overhead~\cite{amazon-tco, crankshaw2019design}.

Most importantly, having SQL and ML running in the same system will unlock co-optimization opportunities and reduce execution costs without affecting accuracy. We summarize the co-optimization techniques into four categories: 


\noindent
 \textbf{(O1) Relational algebra optimization.} An inference query may involve AI/ML-based \texttt{filter}s that invoke AI/ML models on the input data to return a Boolean value, such as \texttt{predict(movieFeatures)}\texttt{=True}. Based on the estimated costs of the AI/ML models and the selectivity of the predicates, the system can push down such \texttt{filter}s through \texttt{join}, without knowing the internal logic of the AI/ML models to avoid redundant computations. Similar techniques include pushing down AI/ML-based \texttt{project} operators, reordering two consecutive customized \texttt{filter}s, and merging multiple consecutive AI/ML-based \texttt{filter}/\texttt{project}s into one. \textcolor{black}{User-Defined Function (UDF)}-centric systems~\cite{kakkar2023eva, hellerstein2012madlib} and systems with extended operators~\cite{gaussml} support some of these techniques. 

\noindent
\textbf{(O2) Factorized inference.} An AI workflow (e.g., a two-tower recommendation model~\cite{yang2020mixed}) may process features that are joined from multiple tables. These workflows/functions can often be factorized into multiple independent functions, each processing features from a separate table. Then, each factorized computation can be pushed down through the \texttt{join} to its corresponding table to avoid redundant computations. The idea was extended from several works~\cite{schleich2016learning, factorize-la, li2019enabling, chowdhury2025inferf} in the context of factorized ML~\cite{schleich2016learning, factorize-lmfao}. 

\noindent
\textbf{(O3) Tensor-relational transformation.} \textcolor{black}{
Model parameters (e.g., weight matrices or decision trees) can be stored as relational tables, and inference can be rewritten into relational operators such as \texttt{join}/\texttt{project}/\texttt{aggregate} over these tables. This makes inference transparent to the database engine and allows it to scale via buffer-pool management and pipelined, parallel execution, even when model parameters are too large to fit in memory
}
~\cite{DBLP:journals/pvldb/YuanJZTBJ21, jankov12declarative, zou2018plinycompute, zou2021lachesis, zou2019pangea, zou2020architecture}. 

\noindent
\textbf{(O4) Data-model and cross-library optimization techniques.} Seeing AI/ML as a white box, there are more cross-optimization techniques, e.g., fusing AI/ML operators from different libraries~\cite{palkar2017weld, palkar2018evaluating},  
 pruning nodes from decision forest models based on the relational predicates applied to the data to be inferred, and pruning attributes from datasets based on model logic~\cite{park2022end}. Another co-optimization technique in this category is to replace ML operators on dense tensors with operators optimized for sparse tensors, assuming that the database is aware that the input data is sparse~\cite{schleich2023optimizing}. 

\noindent
\textbf{The Problem and the Challenges.} As detailed in Sec.~\ref{sec:related-works}, most of the existing DB-for-AI systems~\cite{gaussml, kakkar2023eva, meng2016mllib, jankov12declarative, del2021transforming, park2022end, 
factorize-la, 
hellerstein2012madlib, postgresml, factorize-lmfao, DBLP:journals/pvldb/YuanJZTBJ21} \textit{cannot} adaptively exploit diverse optimization techniques in O1-O4 for different sub-computations, and thus they will miss the optimization opportunities brought by combining techniques from more than one category, leading to sub-optimal query latency.  The research challenges include:

\noindent
\textbf{Challenge 1. Difficulties in unifying all co-optimization techniques in one system.}
Each category of co-optimization techniques relies on a different data/computational abstraction, cost model, and transformation logic. Relational algebra optimizations (O1) rely on classical query planning rules and cost models based on the statistics and selectivity of opaque UDFs. Factorizing an ML function across joins (O2) requires analyzing the internal structure of the UDFs and the data schema together. Tensor-relational rewrites (O3) not only require analyzing the internal logic of UDFs but also converting tensor data into collections of tensor blocks for relational processing. Data-model and cross-library optimization (O4) demand reasoning about both data layout and model internals. Existing intermediate representations and systems typically support only limited combinations of these abstractions and lack mechanisms to express all required cross-level optimizations in a coherent, cost-aware way.
%
\textbf{Challenge 2. Massive Search Space.} 
Our three-level \textcolor{black}{Intermediate Representations (IR)} has enabled full co-optimization opportunities, which enlarged the search space. Supposing there are $n$ atomic ML functions, each function has two options, converting to relational operators or not, which leads to $2^n$ optimization plans in the O3 category. Then, applying O3 techniques brings new relational operators, which will expand O1's search space. 
In O4, depending on how $n$ consecutive atomic ML functions should be fused, we will have $2^{n-1}$ plans. \textcolor{black}{If an atomic ML function is not fused with other functions, it may bring new optimization opportunities in O2 and O3.} 
It is essential to explore IR-optimizer co-design to effectively manage this expanded search space.

\noindent

To address these challenges, we present \CactusDB, a novel UDF-centric system with:


\noindent
(1) \textbf{A Three-Level IR (Sec.~\ref{sec:ir}).} 
To address the first challenge, we advocate a three-level IR 
on top of the approach combining relational and linear algebra, to capture computation at different layers.
The top level consists of relational operators nested with opaque user-defined expressions.
At the middle level, each opaque expression at the top level is lowered to analyzable tree structures, in which each node is either an expression operator or an opaque expression.
At the bottom level, each opaque expression from the middle-level IR is (optionally) described as a computational graph of ML or UDF functions.
We leverage the underlying Velox system for query/expression parsing, vectorization, and physical optimization. 


\noindent
(2) \textbf{A Novel Reusable MCTS Optimizer (Sec.~\ref{sec:optimization}).} Today's cost-based query optimizers increasingly adopt the MCTS for efficiently exploring large search spaces,
balancing exploitation and exploration via the Upper Confidence Bound (UCB) mechanism~\cite{browne2012survey}, and integrating with diverse models. 
However, existing MCTS-based query optimizers~\cite{zhou2021learned, yu2022cost, bai2023maliva, sikdar2020monsoon} need to build a search tree from scratch for each query, as each has distinct states (query plans) and action spaces (rewrite rules), making reuse of MCTS tree challenging.
We address this by combining query embeddings with MCTS to form a universal state space.
Our embedding design aligns with the three-level IR, while the action space consists of co-optimization rules (O1-O4).

\noindent
(3) \textbf{Implementation and A Comprehensive Evaluation (Sec.~\ref{sec:evaluation}).} We implemented \CactusDB\footnote{Open source at: \url{https://github.com/asu-cactus/cactusdb}} on top of Meta's open-sourced Velox database engine~\cite{pedreira2022velox}. 
We further evaluated our system on several benchmarks, including (1) realistic recommendation queries on MovieLens dataset~\cite{harper2015movielens, sarwat2017database} (2) TPCx-AI workloads~\cite{brucke2023tpcx}, and (3) analytics workloads on three real-world datasets - Credit Card~\cite{kaggle-fraud}, Expedia\textcolor{black}{~\cite{projecthamlet2021}}, and Flights\textcolor{black}{~\cite{projecthamlet2021}}. For the MovieLens and TPCx-AI workloads, we designed ten representative query templates for each by extracting and generalizing patterns from the most popular Kaggle notebooks. We then sampled \textcolor{black}{$2{,}000$} queries based on these templates for evaluation. We observed up to $441 \times$ speedup using \CactusDB compared to baseline systems.

\section{Background}
\label{sec:background}

\subsection{SQL-ML Co-Optimization}
\label{sec:co-optimization-rules}

First, without loss of generality, we assume that pure relational algebra optimization, such as join reorder and the push-down of simple predicates and projections that do not involve ML, is applied before performing co-optimization, and typical database physical optimization, e.g., join algorithm selection, is applied after co-optimization.

Then, we categorize (non-approximate) SQL-ML co-optimization techniques into four main classes, O1–O4, each defined by its distinct data abstraction (particularly how model parameters are represented) and computation abstraction. 


\vspace{3pt}
\noindent
{\color{black}
\textbf{O1. Relational Algebra Optimization}~\cite{evadb_optimization, gaussml, postgresml, hellerstein2012madlib}.
O1 performs optimization at the relational algebra level.
The \emph{computation abstraction} consists of standard relational operators such as
\texttt{filter}, \texttt{project}, \texttt{join}, \texttt{crossJoin}, and \texttt{aggregate}.
AI/ML inference is encapsulated within user-defined expressions and treated as opaque functions.
The \emph{data abstraction} consists solely of relations; model parameters (e.g., weight tensors)
are not exposed to the optimizer.
Despite this opacity, the optimizer can reduce the \emph{number} and \emph{placement} of AI/ML
invocations by rewriting the relational query plan.
Representative co-optimization rules in this category (R1-1 to R1-4) are summarized below. Rules in this category preserve relational algebra equivalence provided that AI/ML models used within UDFs are deterministic. 

\begin{shaded}{
\smaller
\noindent
\textbf{R1-1 \texttt{filter} Reorder.}
Reordering \texttt{filter} operators, including those containing user-defined expressions,
preserves query equivalence. Ordering filters by selectivity reduces intermediate result sizes.

\noindent
\textbf{R1-2 \texttt{filter} Pushdown.}
Pushes a \texttt{filter} down through a \texttt{join} to avoid redundant inference over high-cardinality
join outputs that usually contain repeated tuples.

\noindent
\textbf{R1-3 \texttt{project} Pushdown.}
Pushes \texttt{project} operators, including those invoking ML inference, below \texttt{join}s
to eliminate unnecessary computation.

\noindent
\textbf{R1-4 Operator Merge/Split.}
Multiple \texttt{filter} or \texttt{project} operators over the same relation can be merged into
a single operator; splitting is the inverse transformation.
}
\end{shaded}

}

{\color{black}
\vspace{3pt}
\noindent
\textbf{O2. Factorized Inference}
(derived from factorized ML~\cite{factorize-la, li2019enabling, factorize-lmfao, chowdhury2025inferf}).
O2 extends relational optimization by exposing a richer computation abstraction that includes
relational operators, linear algebra primitives, and a restricted set of factorizable ML operators
(e.g., vector--matrix multiplication).
The \emph{data abstraction} explicitly exposes model parameters as factorizable objects, enabling
the optimizer to reason about shared computation across joined relations.
Representative co-optimization techniques are illustrated below.

\begin{shaded}{
\smaller
\noindent
\textbf{R2-1 Linear Algebra Operator Factorization}~\cite{factorize-la, li2019enabling, factorize-lmfao}.
Let $S(Y, \boldsymbol{X}_S, FK)$ and $R(\underline{RID}, \boldsymbol{X}_R)$ be two relations with feature
vectors $\boldsymbol{X}_S$ and $\boldsymbol{X}_R$.
Joining on $S.FK = R.RID$ produces $T(Y, \boldsymbol{X})$, where
$\boldsymbol{X} = [\boldsymbol{X}_S, \boldsymbol{X}_R]$.
Consider a linear inference operation that computes $w^\top \boldsymbol{x}
\top$ for each
$\boldsymbol{x} \in T.\boldsymbol{X}$.
Since the join replicates feature vectors from $S$ and $R$ across many output tuples,
naively evaluating $w^\top \boldsymbol{x}^\top$ on $T$ incurs redundant computation.
By partitioning the weight matrix as $w = [w_S, w_R]$, the computation can be rewritten as
$w_S^\top \boldsymbol{x}_S^\top + w_R^\top \boldsymbol{x}_R^\top$.
This factorization enables pushing $w_S^\top \boldsymbol{x}_S^\top$ to $S$ and
$w_R^\top \boldsymbol{x}_R^\top$ to $R$ prior to the join, and combining the results by addition
after the join, thereby eliminating repeated matrix multiplications.
An example is shown in Fig.~\ref{fig:factorization-example}.
}
\end{shaded}

Beyond linear models, equivalent factorization strategies have been developed for linear
regression, \textit{k}-means clustering, decision trees, and gradient-boosted trees
(e.g., Morpheus~\cite{li2019enabling}, LMFAO~\cite{factorize-lmfao}, JoinBoost~\cite{factorize-joinboost}, InferF~\cite{chowdhury2025inferf}),
with formal equivalence proofs provided in these works.

}

{\color{black}
\begin{figure}[h]
\centering
\includegraphics[width=0.485\textwidth]{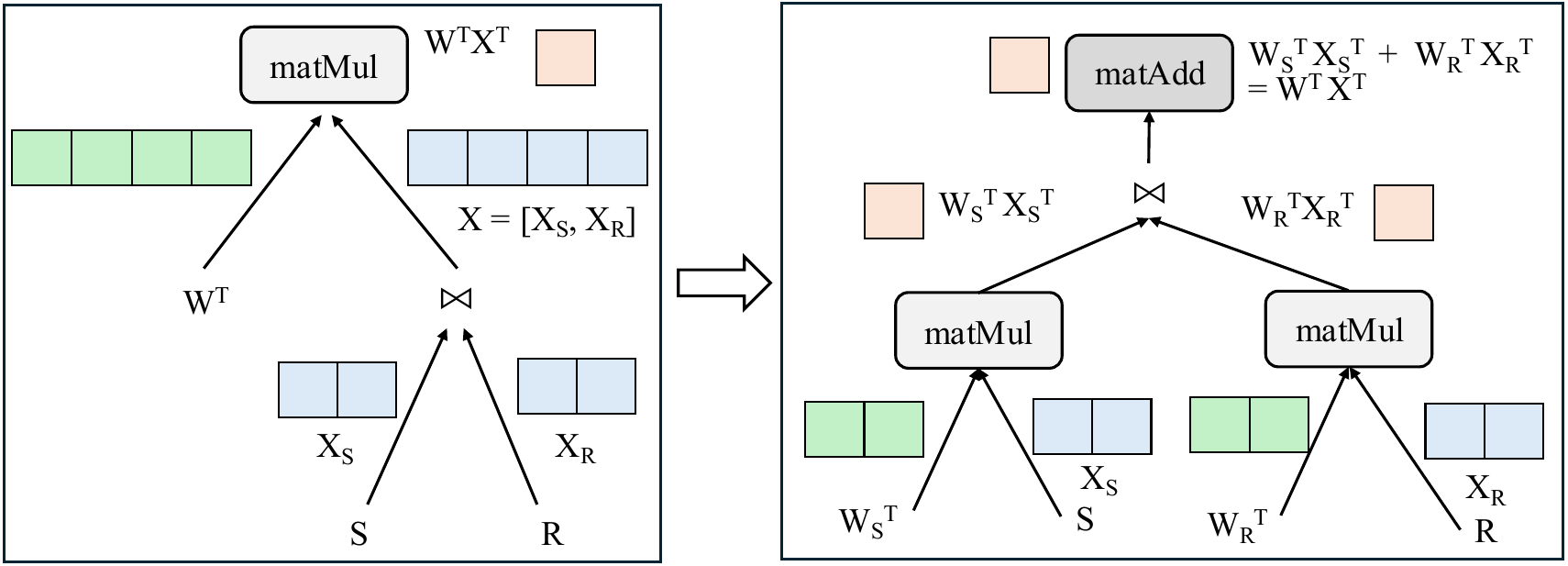}
\caption{\label{fig:factorization-example} \smaller
\textcolor{black}{Example Factorization of Matrix Multiplication (R2-1)}}
\end{figure}
}

\vspace{3pt}
\noindent
 \textbf{O3. Tensor-Relational Transformation.}
 The \underline{computation ab-} \underline{straction}: all AI/ML models are transformed into a limited set of tensor-relational operators~\cite{DBLP:journals/pvldb/YuanJZTBJ21}, such as \texttt{aggregate}, \texttt{crossJoin}, \texttt{filter}, \texttt{project}, \texttt{rekey}, \texttt{tile}, \texttt{concat}, all processing tensor relations. 
In the \underline{data abstraction}, the model parameters must be transformed into tensor relations. 
\underline{Example co-optimization techniques} include:

\begin{shaded}
\smaller
\noindent
\textbf{R3-1} \textbf{Linear algebra operators transformed to relational operators}~\cite{DBLP:journals/pvldb/YuanJZTBJ21, luo2018scalable, boehm2016systemml, DBLP:journals/pvldb/ZhouCDMYZZ22, zou2020watson}. 
{\color{black}{Consider computing $Y = xW^{\top}$ for $N$ input feature vectors, where $x$ represents a row from the input table (i.e., a feature vector), and the weight matrix $W$ is too large to fit in memory.
R3-1 stores $W$ as a \emph{tensor relation} of column blocks,
$\widetilde{W}(\textit{colId}, \textit{wTile})$, and rewrites inference into a simple relational pipeline:
\texttt{crossJoin}$\rightarrow$\texttt{project}, producing block outputs
$B(\textit{rowId}, \textit{colId}, \textit{yBlock})$.
At runtime, the DB scans $\widetilde{W}$ one tile at a time.
For each $\textit{wTile}$ loaded into the buffer pool, it computes the corresponding output block
$\textit{yBlock} := x \cdot \textit{wTile}^{\top}$ for all rows, emits $B$, and then loads the next block. 
Finally, it assembles each output vector by concatenating $\textit{yBlock}$ blocks per \textit{rowId}. The process is illustrated in Fig.~\ref{fig:O3-examples}.}}
{\color{black}
As described in Tensor Relational Algebra (TRA)~\cite{DBLP:journals/pvldb/YuanJZTBJ21}, other linear algebra computations can also be represented in relational algebra. 

\noindent
\textbf{R3-2} \textbf{Decision Forest to relational operators}~\cite{guan2023comparison}. A decision forest model, like \texttt{XGBoost} or \texttt{lightGBM}, is stored as $DF(\underline{treeId}: INT, trees: DecisionTreeObject)$. Applying the model $DF$ to each $x \in T.\boldsymbol{X}$, could be transformed into relational processing by first performing a \texttt{crossJoin} of $T$ and $DF$. Then a \texttt{project} is applied to each joined pair of $(x \in T.\boldsymbol{X}$, $t \in DF.trees)$ to run $t.predict(x)$. The prediction results are aggregated to obtain the final prediction, e.g., via majority voting or a sigmoid function.}
\end{shaded}

{\color{black}
\begin{figure}[h]
\centering
\includegraphics[width=0.45\textwidth]{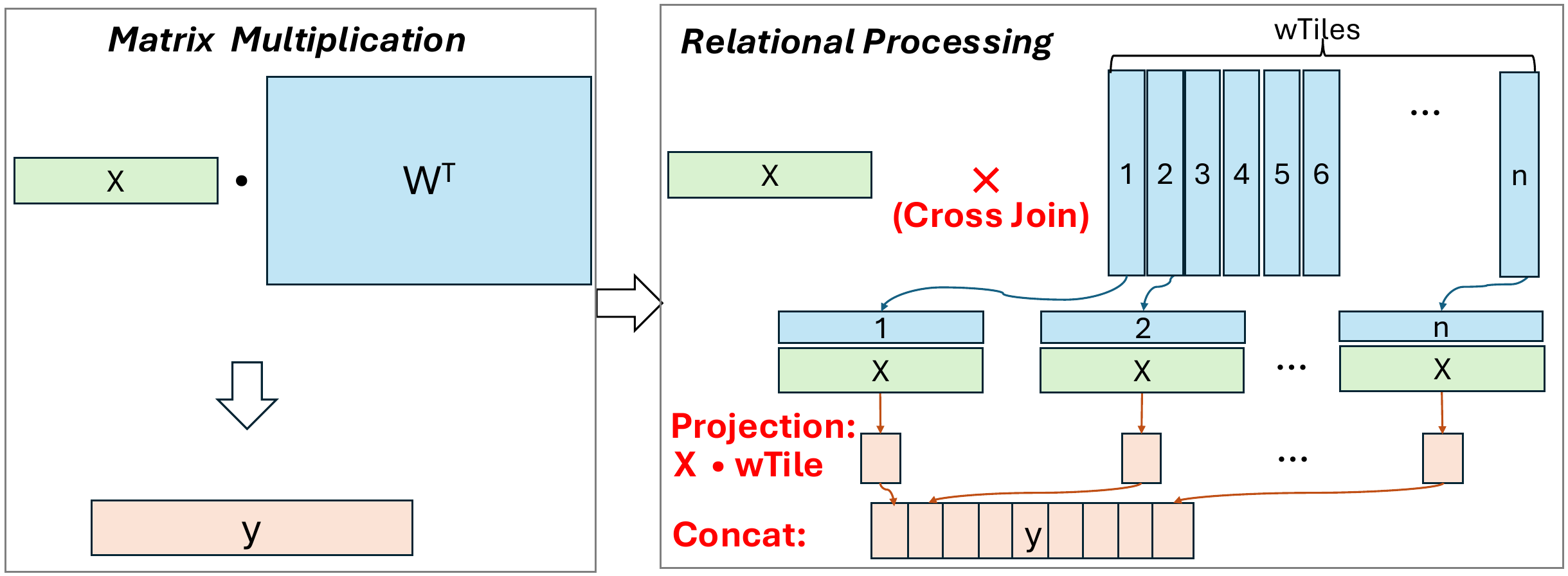}
\caption{\label{fig:O3-examples} \small
\textcolor{black}{Example of \texttt{MatMul} to Relational Processing (R3-1)}}
\end{figure}
}

The O3 transformations scale to AI/ML models that involve a large volume of parameters, by blocking the parameters so that each time only a few blocks need to be loaded into the memory (via buffer pool management). \textcolor{black}{The formal proof for the equivalence of transformation rules in O3 was provided in prior work~\cite{DBLP:journals/pvldb/YuanJZTBJ21}.}

\vspace{3pt}
\noindent
{\color{black}
\textbf{O4. Data--Model Cross Optimization}~\cite{park2022end, palkar2017weld, kunft2019intermediate, shaikhha2022functional}.
O4 removes the abstraction boundary between data processing and model execution.
The \emph{computation abstraction} spans relational operators, linear algebra primitives,
and AI/ML operators, while the \emph{data abstraction} jointly includes relations and tensor
objects (e.g., model parameters and intermediate activations).
By explicitly exposing AI/ML operators and tensor data, the optimizer can jointly reason
about data access patterns, operator structure, and execution strategies across the
data--model boundary.
Representative optimization techniques are summarized below.

\begin{shaded}
\smaller
\noindent
\textbf{R4-1 Operator Fusion and Split.}
Consecutive AI/ML operators can be fused into a single composite operator to reduce
materialization and invocation overhead.
For example, a sequence of matrix multiplication, matrix addition, and activation
(e.g., \texttt{matMul} $\rightarrow$ \texttt{matAdd} $\rightarrow$ \texttt{sigmoid})
can be replaced with a single dense-layer operator from a deep learning library.
Conversely, a monolithic UDF can be split into finer-grained operators, enabling
selective application of factorization or parameter-aware optimizations
(e.g., O2 and O3) to sub-computations.

\vspace{2pt}
\noindent
\textbf{R4-2 AI/ML Library Replacement.}
The physical implementation of an AI/ML operator can be substituted with alternative
library backends based on hardware characteristics or data properties.
For instance, \texttt{matMul} may be executed using CPU, GPU, sparse, or dense kernels.
Similarly, a \texttt{conv2D} operator can be rewritten as \texttt{matMul} via spatial
reorganization of inputs and filters~\cite{vasudevan2017parallel, Conv-spatial-rewrite, spatial-rewrite}.
Other techniques in this category include data-induced model optimization~\cite{park2022end}
and inference batch size optimization~\cite{zhang2024imbridge}.
\end{shaded}

The correctness of the above rules (i.e., transformation equivalence) relies on the functional equivalence of alternative
physical implementations of the same logical operators.
In practice, systems ensure equivalence through library guarantees and empirical validation.
}

\textcolor{black}{In this paper, we advocate supporting all categories of optimization techniques within a unified database system. Such a design allows different sub-computations to operate under different data abstractions and co-optimization regimes, thereby enabling optimization opportunities that are inaccessible when techniques are applied in isolation. For example, after applying operator splitting (R4-1) to decompose a dense-layer UDF into finer-grained operators, linear algebra primitives such as \texttt{matMul} become explicitly visible to the optimizer, making them eligible for factorization- and parameter-aware optimizations (e.g., R2-1 and R3-1).}

\vspace{-5pt}
\section{Three-Level IR}
\label{sec:ir}

In this section, we introduce our data and computation abstractions and describe our three-level IR design and usage.

\subsection{Flexible Data Abstraction}
\label{sec:data-abstraction}

We abstract features and model variables as follows:

%
\noindent
\textbf{Feature Vectors} are extracted from relations and stored alongside other attributes, denoted as $F(A_1: Type_1, \dots, A_l: Type_l, \boldsymbol{V}: \mathbf{vec} \in \mathbb{R}^d)$, where $A_i$ is an attribute of type $Type_i$, and $\boldsymbol{V}$ represents the stored feature vectors. If $\boldsymbol{V}$ concatenates $m$ vectors $V_1,\dots,V_m$, where $V_1,\dots,V_m$ are from datasets $D_1,\dots, D_m$, respectively, we define $\boldsymbol{V}\equiv[V_1,\dots,V_m]$. 
\noindent
\textbf{Model Parameters} are variables of ML functions, which are created when models are loaded into \CactusDB. A variable can be a tensor (deep learning), a tree collection (decision forest), or other types of parameters.  A variable could be stored as a relation. For example, a $d$-dimensional weight tensor $W_i$ can be stored as a relation $P_i(dim_1: INT,\dots, dim_d:INT, block: \mathbf{vec} \in \mathbb{R}^{k_1\times\dots\times k_d})$, where each tuple represents a $d$-dimensional tensor block indexed by $dim_1,\dots,dim_d$, with the block data of shape $k_1\times\dots\times k_d$ flattened as an array in the column $block$. Similarly, a decision forest can be stored as a relation $DF(TreeId: INT, TreeObj: TreeType)$, where each tuple represents a tree object indexed by $TreeId$. 
%
To enable efficient conversion, \CactusDB selectively materializes model variables as relations during loading if their size exceeds a threshold (typically half the available memory).

\subsection{Three-Level Computation Abstraction}
\label{sec:computation}

\CactusDB supports three types of operators:

\noindent
\textbf{Relational Operators.}
\CactusDB supports most relational operators such as \texttt{filter}, \texttt{project}, \texttt{join}, \texttt{crossJoin}, \texttt{union}, \texttt{aggregate}, and \texttt{expand} (i.e., similar to \texttt{flatmap}). Each such operator processes a relation, such as $F$, $P_i$, $DF$.

\noindent
\textbf{Expression.} Most relational operators should be customized by expressions,  which are in the form of analyzable tree structures. A tree node can either be an opaque expression or an expression operator, such as arithmetic (+, -, *, /), comparison (>, <, =, $\neq$), bit (e.g., $\wedge$, $\vee$), relation (e.g., AND/OR), conditional (e.g., IF/THEN/ELSE and SWITCH), and function call (e.g., CALLFUNC) operators. Each expression operates on one or multiple columns, e.g., $A_i, \boldsymbol{V}$, $\dim_j, \boldsymbol{block}$.








\noindent
\textbf{ML Functions.} In \CactusDB, we have implemented a set of ML functions. An ML function is a specialized Velox UDF.
%
Each ML function class has an \texttt{apply} method, which takes a batch of vectors or tensor blocks as input to facilitate the use of SIMD instructions and reduce virtual function call overheads. Although this work focuses on CPU-based database systems, GPU acceleration can also be enabled for many ML functions by passing a flag to the \texttt{apply} method. 
The supported functions include: (1) \textbf{Atomic ML Functions} such as \texttt{matMul}, \texttt{matAdd}, \texttt{conv}, \texttt{softmax}, \texttt{argmax}, \texttt{pooling}, \texttt{relu}, \texttt{batchNormalization}, \texttt{cosSim} (cosine similarity),  \texttt{concat}, \texttt{drop}\texttt{out}, \texttt{embed} (embedding lookup), \texttt{decision} \texttt{Tree}, \texttt{standard/m}\texttt{in/maxScalarization}, \texttt{seqEncode} (i.e., encoding a sequence of tokens into an embedding vector), and \texttt{binarization}; (2) \textbf{High-level ML Functions} such as
 feed-forward neural network model (\texttt{ffnn}), convolutional neural network (\texttt{cnn}), decision forest model (\texttt{xgboost/lightgbm}), Huggingface inference endpoints (\texttt{huggingface}), and ChatGPT APIs (\texttt{llm}). These functions take one or more vectors/tensor objects as input, e.g., $W_i \in \boldsymbol{W}$, $\boldsymbol{vec}\in\boldsymbol{V}$,$\boldsymbol{vec}\in\boldsymbol{block}$ as described in Sec.~\ref{sec:data-abstraction}.




Importantly, a user can register customized atomic and high-level functions in addition to these built-in ML functions.


\vspace{-10pt}
\subsection{IR Overview}
To maximize co-optimization opportunities by enabling each sub-computation to select its optimal computation and data abstraction, we propose a novel three-level IR:

\noindent
\textbf{Top-level IR: Relational Algebra.}
In the top-level IR, every node is a relational operator, which is customized by an expression opaque at this level. Each edge represents a data flow from the source node to the destination node.

\noindent
\textbf{Middle-level IR: Expression Tree.} An opaque expression from the top-level IR is lowered to the middle-level IR as an expression tree, where each node is either an expression operator (see Sec.~\ref{sec:computation}) or an opaque sub-expression.  
%


\vspace{3pt}
\noindent
\textbf{Bottom-level IR: ML Computation Graph.}
\label{sec:bottom}
An opaque sub-expression in the middle-level IR is lowered to the bottom-level IR as a computation graph that is understandable to the query optimizer. In the bottom-level IR, each node represents an atomic ML function (see Sec.~\ref{sec:computation}), while each edge represents a tensor output from the source node and input to the destination node. From the bottom-level IR, each atomic ML function's details, such as input tensor shapes, the shapes of model weight matrices, etc., can be retrieved by the query optimizer from the ML function instance through pre-defined interfaces. 

Our IR design allows a query optimizer to rewrite more than one level of IRs at the same time to support any co-optimization techniques from O1-O4 in arbitrary ordering for any sub-computation, maximizing co-optimization space, as shown in the next section\textcolor{black}{.}

{\color{black}
\subsection{Workflow: from a Query to Our IR} 
\label{sec:ir-workflow}
As illustrated in Fig.~\ref{fig:IR-workflow}, at Step 1, a user loads a two-tower recommendation model, which is a popular type of recommendation models~\cite{yang2020mixed, li2022inttower}. In our example, the model consists of a user tower and a movie tower. The two towers encode user features and movie features into vectors, respectively. The cosine similarity between the user vector and the movie vector is then computed to decide whether the movie should be recommended to the user. During the model loading process~\cite{cactusdb-model-loading}, all ML functions required by the model are developed if needed (can build on existing AI/ML libraries), customized, registered, and used to compose a computational graph. At Step 2,  the computational graph is registered as the \texttt{twoTowerModel} expression.  At Step 3, a query is issued that applies the \texttt{twoTowerModel} expression to compute a recommendation score for each pair of preprocessed user and movie features. At Step 4, the query is translated into a three-level IR, where the top \texttt{project3} node in the top-level IR points to a middle-level IR that only contains the opaque \texttt{twoTowerModel} expression, which links to the detailed bottom-level IR registered at Step 3. 

\begin{figure}[h]
\centering

\includegraphics[width=0.485\textwidth]{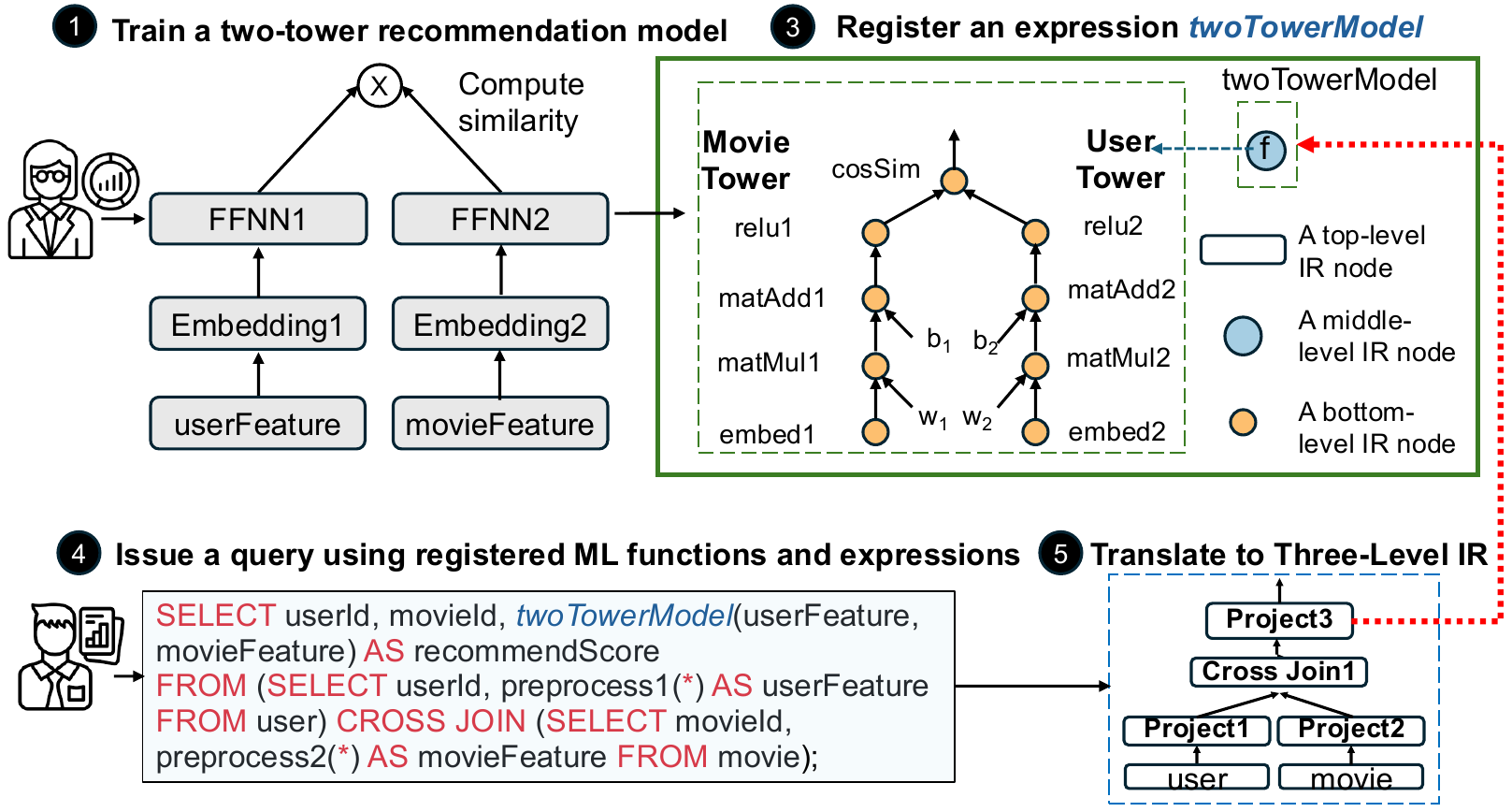}

\caption{\label{fig:IR-workflow} \small
 \textcolor{black}{Develop and Translate an Inference Query to Our IR. }}
\end{figure}
}

\subsection{Examples of Using Our IR}
\label{sec:ir-examples}
\begin{figure}[h]
\centering
\includegraphics[width=0.5\textwidth]{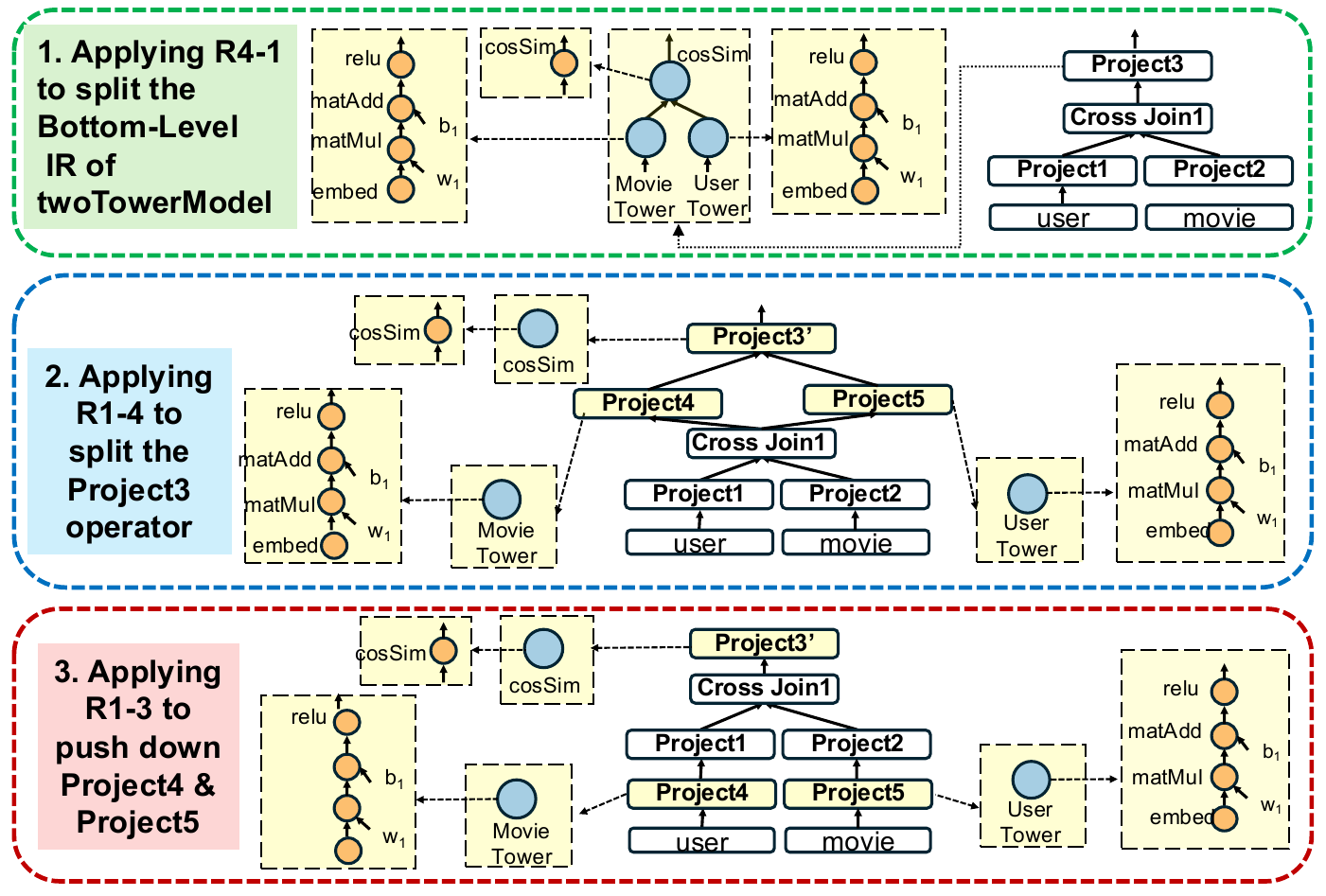}
\caption{\label{fig:running-examples} \small
\textcolor{black}{Example Transformations of our three-level IR.}}
\end{figure}

{\color{black}
Fig.~\ref{fig:running-examples} presents three transformations of the 3-level IR from our running example in Fig.~\ref{fig:IR-workflow}.}

As shown in Fig.~\ref{fig:running-examples}-1, the optimizer first splits the \texttt{twoTowerModel} expression into three parts: the user tower, the movie tower, and an operator that computes their cosine similarity (See R4-1 in Sec.~\ref{sec:co-optimization-rules}). In Fig.~\ref{fig:running-examples}-2, the node \texttt{Project3} in the top-level IR is split into \texttt{Project3'} corresponding to cosine similarity, \texttt{Project4} corresponding to movie tower, and \texttt{Project5} corresponding to user tower (See R1-4 in Sec.~\ref{sec:co-optimization-rules}). Then, in Fig.~\ref{fig:running-examples}-3,  \texttt{Project4} and \texttt{Project5} are pushed down to avoid redundant computations (See R1-3 in Sec.~\ref{sec:co-optimization-rules}).

\vspace{-5pt}
\section{Query Optimization}
\label{sec:optimization}


In this section, we will describe our unique query optimization process co-designed with our three-level IR representation. The goal is to address the search space challenge via a novel reusable MCTS model based on query embedding techniques.

\vspace{-5pt}
\subsection{MCTS Preliminary}
\label{sec:mcts-vanila}
MCTS has become a popular tool for query optimization~\cite{zhou2021learned, sikdar2020monsoon}. That is because MCTS can efficiently explore a large search space, e.g., various rewrite ordering, by balancing exploration and exploitation, guaranteeing asymptotic convergence to the optimal decision once combined node selection strategies such as UCB~\cite{chang2005adaptive}. A common abstraction of MCTS for query optimization includes the following components:

\noindent
\textbf{States}, i.e., a node in the MCTS tree, represents a logical query plan of the SQL query to be optimized and the input datasets. 

\noindent
\textbf{Node Utility Function.} Every state is associated with a utility representing the cumulative reward of the state and indicating the probability that the state is in the optimal path. 

\noindent
\textbf{Actions.} Usually, an action, i.e., an edge in the MCTS tree, refers to a rewriting or equivalent transformation of the current plan, transitioning it to a new plan that produces equivalent results. The applicable rewriting and transformation rules defines the action space. 

\noindent
\textbf{Runtime Search Process.} The MCTS algorithm starts with the root node, the initial query plan. 
At each step, it checks whether the current state has been expanded, i.e., all rewriting rules of the given state have been explored. If so, the MCTS selects an action based on UCB (Line $1$ of Alg. \ref{alg:mcts-select}), where $r_i$ is the utility of $node_i$ (the accumulative reward of $node_i$), $n_i$ is the number of total visits of the $node_i$, $N_i$ is the total number of visits of the parent node of $node_i$, and $c$ is a constant for balancing exploration ($\frac{r_i}{n_i}$) and exploitation ($\sqrt{\frac{\ln{N_i}}{n_i}}$). 
If the current node is not fully expanded, it will expand the node by randomly selecting one unexplored action (see Alg.~\ref{alg:mcts-expand}). Then, it performs a simulation, also called a rollout, as described in Alg.~\ref{alg:mcts-simulation}. The rollout process randomly chooses an action until it reaches a maximum number of iterations or the terminating state (i.e., no more applicable rules). After rolling out, it will run a back-propagation process to compute rewards and update the utility of each selected node, described in Alg. \ref{alg:mcts-update}. \textcolor{black}{See supplementary material for the overall algorithm.} 

\begin{algorithm}[h]
\footnotesize
\SetKwInOut{Input}{Input}
\Input{
$node$: the current state of the given query}
$selectedNode$ $\gets$ $\underset{node_i \in node.children}{\mathrm{argmax}}{\frac{r_i}{n_i} + c*\sqrt{\frac{\ln{N_i}}{n_i}}}$ \;
$\textbf{return}$ $selectedNode$;
\caption{$select(node)$}
\label{alg:mcts-select}
\end{algorithm}

\begin{algorithm}[h]
\footnotesize
\SetKwInOut{Input}{Input}
\Input{
$node$: the current state of the given query}
\resizebox{0.89\linewidth}{!}{$unexplored \gets \{a \in node.actionSpace \mid a.explored == False\}$}\;
\textcolor{black}{$selectedAction$ $\gets$ $randomSelect(unexplored)$ \;}
$selectedAction.explored$ $\gets$ $True$ \;
$node'$ $\gets$ $node.takeAction(selectedAction)$ \;
$node.children$ $\gets$ $node.children \cup \{node'\}$\;
$\textbf{return}$ $node'$;
\caption{$expand(node)$}
\label{alg:mcts-expand}
\end{algorithm}


\begin{algorithm}[h]
\footnotesize
\SetKwInOut{Input}{Input}
\Input{
$node$: the node where simulation starts}
\While {node.isTerminateState() == False}{
    $selectedAction$ $\gets$ $RandomSelect(node.actionSpace)$ \;
    $node$ $\gets$ $node.takeAction(selectedAction)$ \;
}
$\textbf{return}$ $node$;
\caption{$rollout(node)$}
\label{alg:mcts-simulation}
\end{algorithm}

\begin{algorithm}[h]
\footnotesize
\SetKwInOut{Input}{Input}
\Input{$node$: the node where simulation start; $reward$: the reward value obtained from reward function} 
\While {node != NULL}{
    $n_i$ $\gets$ $n_i+1$ \; 
    \textcolor{black}{$r_i$ $\gets$ $r_i + reward$ \;}
    $node$ $\gets$ $node.parent$ \;
}
$\textbf{return}$;
\caption{$backpropagate(node, reward)$}
\label{alg:mcts-update}
\end{algorithm}

\noindent
\textbf{Shortcomings of the naive MCTS optimization.} As a runtime query planner, MCTS constructs a search tree from scratch for every query, making optimization \textcolor{black}{inefficient}.
\textcolor{black}{In addition, simply caching MCTS states for reuse cannot fully resolve the problem, because in practice, repeated queries often use different parameters~\cite{wu2024stage}, which have unique states and action spaces, preventing tree reuse even among similar queries.}

%
Therefore, the naive MCTS model cannot exploit such similarity 
to reduce query optimization latency.

\subsection{Reusable MCTS based on Query Embedding}\label{sec:universal-optimizer}

\begin{figure*}[h]
\centering
\includegraphics[width=0.99\textwidth]{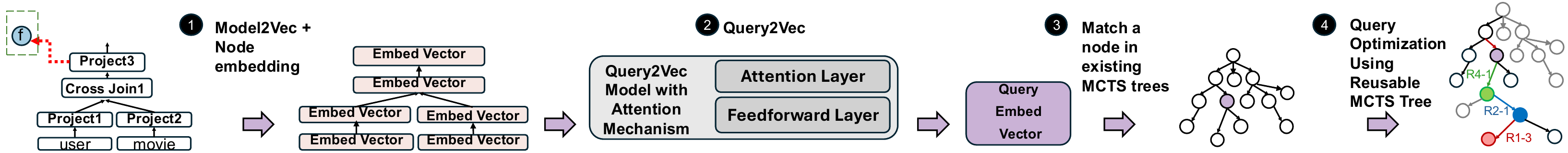}

\caption{\label{fig:optimizer-workflow} \small
 \textcolor{black}{\CactusDB's Query Optimization Workflow. }}
 \vspace{-15pt}
\end{figure*}

Recent Query2Vec techniques encode SQL queries into fixed-length vectors for downstream tasks such as query optimization (e.g., BAO~\cite{marcus2021bao}, NEO~\cite{marcus12neo}), cost estimation (e.g., E2E-Cost~\cite{sun13end}, RTOS~\cite{yu2020reinforcement}), and join/index selection (e.g., ReJOIN~\cite{marcus2018deep}, AIMeetsAI~\cite{ding2019ai}). More recently, QueryFormer~\cite{zhao2022queryformer} leverages transformers for general-purpose embeddings.
We build on this line of work by {{encoding logical query plans into embedding vectors to enable MCTS to generalize across queries and improve optimization efficiency}}. However, existing methods are faced with the following challenges.

\noindent
\textbf{Challenge-1.} \textit{Existing query2vec models did not consider inference queries in three-level IR forms, and they cannot capture expressions and ML functions.}

\vspace{3pt}
\noindent
\textbf{Challenge-2.} \textit{While the query embedding can be used to unify the state space for different queries, the action space remains query-specific. It is difficult to integrate Query2Vec with MCTS to navigate the full co-optimization space.}

\vspace{3pt}
To address the challenges, we offer a novel reusable MCTS framework co-designed with our three-level IR, by extending QueryFormer~\cite{zhao2022queryformer}, the state-of-the-art query embedding technique \textcolor{black}{that is capable of capturing the structural information of the query plan}~\cite{zhao2023comparative}.

\subsubsection{Embedding the Three-level IR}
\label{sec:embedding}
QueryFormer~\cite{zhao2022queryformer}, like other query embedding methods, is designed for tree-based query plans rather than query plans represented in our three-level IR. This is done by encoding each node's attributes, such as \textit{operator type}, \textit{join type}, \textit{input table}, \textit{filter predicate}, \textit{histograms of the related columns}, and \textit{bitmap of samples} into embedding vectors $E_o$, $E_j$, $E_t$, $E_p$, $E_h$, and $E_s$ respectively. These vectors are concatenated to form a fixed-length node vector. Each node vector is further added with a height encoding vector that specifies the node's position in the query graph. The query graph is then abstracted as a sequence of node vectors following the in-order tree traversal. The sequence is sent to a transformer model, which outputs the final embedding vector for the entire query tree. 



{
\color{black}
To address \textbf{Challenge-1}, we adopt a recursive encoding strategy that captures the rich hierarchical structure of the three-level IR. At the bottom level, each opaque expression is encoded into an embedding vector $E_{expr}$ \textcolor{black}{(see Step 1 of Fig.~\ref{fig:optimizer-workflow})}. These are then composed into middle-level IR embeddings, which are used to encode corresponding relational operators. 
Finally, the top-level IR, comprising these embedded relational operators, is processed by a QueryFormer-like model to generate the final query embedding \textcolor{black}{(see Step 2 of Fig.~\ref{fig:optimizer-workflow}).} 

\vspace{3pt}
\noindent
\textbf{Embedding of Bottom-Level IR.} We employ a QueryFormer-like, transformer-based Model2Vec to embed a bottom-level IR into a vector representation in the following approach:
(1) For each opaque expression at a middle-level IR, a \textbf{breadth-first traversal} is applied to its corresponding bottom-level IR (if present), to encode each node (e.g., an ML function) into a vector by concatenating the following components:
\begin{itemize}[leftmargin=*]
    \item \textbf{$E_{mlType}$}: Encodes the ML function type, preserving function diversity. It is generated through an embedding layer, where the parameters are learned during the training process.
    \item \textbf{$E_{mlFlops}$}: Captures the computational cost of the node as the number of floating point operations (FLOPs).
    \item \textbf{$E_{mlDims}$}: Records the shape of the tensor dimensions.
\end{itemize}
The resulting bottom-level IR graph, where each node is represented as a feature vector, is then passed to the Model2Vec to generate the final embedding vector, $E_{expr}$, for the expression. 
}

\noindent
\textbf{Embedding of an Entire Query.} Similar to QueryFormer, for each relational operator in the top-level IR, we first extract feature vectors for $E_o$, $E_j$, $E_t$, $E_p$, $E_h$, and $E_s$. Here, if the operator's middle-level IR contains an opaque expression that maps to a bottom-level IR encoded as $E_{expr}$, $E_i$ is constructed similarly to QueryFormer~\cite{zhao2022queryformer} except that it sees the bottom-level IR's output as an output column encoded by $E_{expr}$.
Then, each feature vector $E_i$ is scaled to a fixed size through a linear layer. The resulting vectors are then concatenated to construct the top-level node embedding vector: 
{\color{black}
\begin{equation}
\setlength{\abovedisplayskip}{5pt}  
\setlength{\belowdisplayskip}{5pt}  
\scriptsize
\label{eq:feature-vector}
V = \parallel LinearLayer_i(E_i), i \in \{o, j, t, p, h, s, expr\}
\vspace{-5pt}
\end{equation}
}
where $\parallel$ is the concatenation. We then map the query plan into a sequence of $m$ vectors ($V_1, ..., V_m$) corresponding to $m$ top-level IR nodes following in-order tree traversal of the top-level IR, which is sent to a QueryFormer-like Query2Vec model~\cite{zhao2022queryformer} to output a final embedding vector for the query plan represented in our three-level IR.
\textcolor{black}{This design enables the estimation of selectivity via two complementary mechanisms. First, for native SQL filters, selectivity features are explicitly encoded via the histogram and filter predicate embeddings. Second, for AI/ML filters, selectivity is learned implicitly by combining the model embedding with the table and column embeddings. During the training process, the model learns the latent relationships between the model's input and output distribution, capturing the predicate's selectivity.}

\vspace{3pt}
 While the architectures of our Model2Vec and Query2Vec models are similar to QueryFormer~\cite{zhao2022queryformer}, we redesign the loss function for two tasks: (1) query embedding for matching a query to an MCTS state; and (2) execution latency prediction for MCTS reward computation. We train these two tasks separately, \textcolor{black}{because joint training of both objectives resulted in gradient interference, degrading the cost estimation accuracy. Decoupling these tasks outperforms the single-model strategy as discussed in Sec.~\ref{sec:optimizer-study}.}
{
\color{black}

\vspace{3pt}
\noindent
\textbf{Task 1. Query Embedding for MCTS State Matching.} 
We combine contrastive learning with Weisfeiler-Lehman (WL) kernel, a widely used and effective method for capturing structural similarity of graphs~\cite{cai2022star}, to map semantically similar expressions and query plans closer in the Model2Vec and Query2Vec embedding spaces while pushing dissimilar ones apart~\cite{gao2021simcse, chen2025automatic}. 
The corresponding contrastive loss is defined in Eq.~\ref{eq:loss-contrastive}, where $N$ is the number of training samples, $\textbf{z}_i$ denotes the embedding vector of the anchor sample, $\textbf{z}_i^{+}$ and $\textbf{z}_i^{-}$ are its positive and negative examples, $sim(\textbf{z}_1, \textbf{z}_2)$ denotes cosine similarity 
$\frac{\textbf{z}_1^{\top}\textbf{z}_2}{||\textbf{z}_1||\cdot||\textbf{z}_2||}$, and $\tau$ is a temperature hyperparameter.

\vspace{-10pt}
\begin{equation}
\setlength{\abovedisplayskip}{5pt}  
\setlength{\belowdisplayskip}{5pt}  
\scriptsize
\label{eq:loss-contrastive}
Loss_{contrastive.} = \frac{1}{N}\sum_{i=1}^{N}\mathcal{L}_i
\vspace{-5pt}
\end{equation}

\begin{equation}
\scriptsize
\label{eq:loss-contrastive-indivi}
\mathcal{L}_i = -\log\frac{\exp(sim(\textbf{z}_i, \textbf{z}_i^{+})/ \tau)}{\exp(sim(\textbf{z}_i, \textbf{z}_i^{-}) / \tau) + \exp(sim(\textbf{z}_i, \textbf{z}_i^{+})/ \tau)}
\end{equation}

{
\color{black}
To train our Model2Vec and Query2Vec models via contrastive learning, we construct positive and negative pairs using structural similarity quantified by the WL subtree kernel~\cite{shervashidze2011weisfeiler}. In each WL iteration, node labels are updated by hashing the current label with the multiset of neighbor labels, capturing rich local structures. Each graph is represented by a feature vector of normalized label frequencies, and similarity is computed via cosine similarity.
In Model2Vec, node labels are initialized by ML function type and FLOPs in the bottom-level IR. Functions of the same type share the same label only if their FLOPs differ within a threshold. In Query2Vec, the initial label for each relational operator encodes operator type, input table, columns, and the WL features of associated expressions.
Although WL features define contrastive pairs, they cannot replace embedding models at runtime, as unseen queries may yield varying label counts and inconsistent feature lengths.

}
%

}

{
\color{black}

\noindent
\textbf{Task 2. Latency Prediction for Cost Estimation.}
We leverage the Query2Vec embeddings to predict the query execution latency, which is used by calculating the reward. This prediction is implemented via a four-layer FFNN that takes the query embedding as input and predicts the estimated latency. The loss function defined in Eq.~\ref{eq:loss-latency}, is based on mean squared error (MSE), where $\hat{y_i}$ and $y_i$ denote the predicted latency and ground-truth latency of the $i$-th sample, respectively.

{
\vspace{-10pt}
\setlength{\abovedisplayskip}{5pt}
 \setlength{\belowdisplayskip}{5pt}
\begin{equation}
\label{eq:loss-latency}
\scriptsize
Loss_{latency} = \frac{1}{N}\sum_{i=1}^{N}(y_i - \hat{y_i})^2
\vspace{-6pt}
\end{equation}
}
}

\vspace{-5pt}
\noindent
We adopt a two-model design. Query2Vec is first trained via contrastive learning for embedding generation, and a separate copy is retrained with an FFNN for latency prediction.



\subsubsection{Reusable MCTS with Configurable Actions}
\label{sec:configuration}
To address the \textbf{Challenge-2}, we proposed a unique MCTS abstraction:

\noindent
\textbf{Embedding-based States.} 
We use our recursive embedding strategy described in the previous sections to convert each logical query plan into a $393$-dimensional embedding vector to represent a state. \textcolor{black}{(This dimensionality is determined by Eq.~\ref{eq:feature-vector}. Following QueryFormer~\cite{zhao2022queryformer}, each of $E_o$, $E_j$, $E_t$, $E_h$, and $E_s$ is $64$-dimensional, The predicate $E_p$ consists of the filter embedding ($64$-dimensional), operator embedding ($8$-dimensional), and a normalized scalar value ($1$-dimensional).)}

\noindent
\textbf{Configurable Actions.} Each action corresponds to a high-level co-optimization rule (Sec.~\ref{sec:background}), making the action space universal across queries. When a rule is selected via UCB, we configure the rule to decide how to apply it to the query. For example, if R3-1 is selected, we must choose a \texttt{matMul} function to convert to relational operators. We first use heuristics, if available, to narrow the candidates, e.g., we select the \texttt{matMul} functions involving the top-$k$ largest tensors. We then identify the optimal candidate that maximizes cost reduction under R3-1 using our embedding-based latency predictor as the cost estimator.
%
The configured rule is then applied to transition the current state to the next state, transforming the query plan. 
This approach allows for a universal action space, addressing Challenge-2.

\begin{algorithm}[t]
\footnotesize
\SetKwInOut{Input}{Input}
\Input {
$query$: a query applied with simple predicate push-down and join reordering;
$M_{Q2V}$: a model embedding $query$ to a vector;
$B\_{iteration}$: the maximum number of iterations;
$mctsTreeSet$: A set of existing MCTS trees;
$nodeIndex$: An index of embedding vectors of existing MCTS tree nodes.
}
    \tcc{\smaller Convert the query into an embedding vector}
    $queryEmbed$ $\gets$ $M_{Q2V}(\text{query})$ \;
    \tcc{\smaller Find nearest neighbor in the MCTS trees}
    \If {nodeIndex.count(queryEmbed)}{
        \tcc{\smaller Start training from the searched node}
        $root$ $\gets$ $nodeIndex.getNearestNeighbor(queryEmbed)$\;
    }
    \Else{
        \tcc{\smaller Otherwise, initialize a new MCTS tree}
        $root$ $\gets$ $TreeNode(queryEmbed)$ \;
        $mctsTreeSet.add(root)$;
    }
    \For {i in range($B\_iteration$)}{
        $node$ $\gets$ $root$ \;
        \While{node.isTerminateState() == False}{
            \If{node.expanded}{
               $node$ $\gets$ $select(node)$ \;
            }
           \Else{
            $node$ $\gets$ $expand(node)$ ;
            $T$ $\gets$ $rollout(node)$ \;
            $\textbf{break}$;
          }
        }
       $reward$ $\gets$ $computeReward(T)$ \;
       $backpropagate(node, reward)$;
    }
    \tcc{\smaller After search, update node indexes}
    $nodeIndex.update(node)$;

$\textbf{return}$;
\caption{Search with Reusable MCTS}
\label{alg:reusable-mcts-search}
\end{algorithm}

\noindent
\textbf{Runtime Search.} A subtree of an MCTS tree may be shared by query plans with similar embedding vectors. As illustrated in Alg.~\ref{alg:reusable-mcts-search}, in the online planning process, a query's default plan will be converted into an embedding vector using our Query2Vec model, denoted as $M_{Q2V}$. \textcolor{black}{As illustrated in Step 3 of Fig.~\ref{fig:optimizer-workflow},} it then performs a nearest neighbor search across the nodes in existing MCTS trees. If there is no matching state with which the cosine similarity is smaller than a threshold, a new MCTS tree needs to be constructed from scratch. Otherwise, it will resume the MCTS search process starting from the matched state\textcolor{black}{, as shown in Step 4 of Fig.~\ref{fig:optimizer-workflow}. Note that in the highlighted sequence of transformation rules that are selected by MCTS, we have R4-1, followed by R2-1, and R1-3, as explained in Sec.~\ref{sec:ir-examples}.} 

\noindent
\textbf{Reward.} Upon reaching a terminate state, $T$, MCTS computes the reward as the overall cost reduction, as the overall cost reduction of the plan, i.e., $reward = cost_{root}-cost_{T}$. This reward is then backpropagated following Alg.~\ref{alg:mcts-update}, while the remaining search process follows the general MCTS procedure 
%
described in Sec.~\ref{sec:mcts-vanila}.

{\color{black}
\noindent
\subsubsection{Error Bound Analysis} 
\label{sec:error-bound}
We model CactusDB’s reusable MCTS-based query optimization as search in a depth-$d$ expectimax tree~\cite{hostetler2014state} rooted at the initial logical plan $s_0$. Let $\mathcal{S}$ denote ground states (logical plans), $\mathcal{A}$ rewrite actions, and $\mathcal{T}$ the set of ground trajectories up to depth $d$. The optimal action-value and value functions over trajectories are defined as follows, where $\mathcal{P}(t, a, t')$ represents that the transition probability of $t$ to $t'$ (one step extension of $t$) for $a\in \mathcal{A}$. We also have $\mathcal{R}(t)$ denoting the reward of selecting trajectory $t \in \mathcal{T}$.
\begin{align}
Q^*(t,a) &= \mathcal{R}(t) + \sum_{t'} \mathcal{P}(t,a,t') V^*(t'), \\
V^*(t) &= \max_{a\in\mathcal{A}} Q^*(t,a).
\end{align}

CactusDB applies a query embedding function $f:\mathcal{S}\rightarrow\mathcal{X}$ to aggregate ground states into abstract states, inducing a partition of trajectories into abstract histories $\mathcal{H}$. Reusable MCTS shares statistics among trajectories within each abstract history, effectively optimizing over the resulting abstract expectimax tree.
Following prior analysis of MCTS with state aggregation~\cite{hostetler2014state}, we assume the embedding $f$ satisfies \emph{$(p,q)$-consistency}: for every abstract history $h\in\mathcal{H}$, there exists an action that is uniformly $p$-near-optimal for all ground trajectories in $h$, and the optimal values of any two trajectories in $h$ differ by at most $q$. Under this assumption, the abstraction-induced error accumulates linearly with search depth.

\noindent
\begin{theorem}[Reusable MCTS error bound under $(p,q)$-consistency]
\label{thm:pq_main}
Let $a^\sharp$ be the action selected at the root by optimizing over the abstract MCTS tree. Then the suboptimality of $a^\sharp$ in the ground MDP is bounded by
\[
\bigl|\max_{a\in\mathcal{A}} Q^*(s_0,a) - Q^*(s_0,a^\sharp)\bigr|
\;\le\; 2d(p+q).
\]
\end{theorem}

\noindent
The full analysis with formal proof is provided in  Appendix~\ref{sec:error-bound-pq} and ~\ref{app:pq_proof}.

\noindent
\textbf{Empirical Error Bound} We additionally derived an empirical error bound using Latin Hypercube Sampling (LHS) to systematically cover the query embedding space. Specifically, we sampled $200$ queries (from queries in Sec.~\ref{sec:synthetic}), and for each query, used our reusable MCTS framework to optimize it with the optimized execution latency $t$. We then performed exhaustive search to obtain the ground-truth optimal latency $t^*$. Across all sampled queries, we observed a mean relative error of $0.9661\%$  with a standard deviation of $0.6355\%$. The corresponding $95\%$ confidence interval ranges from $0.8771\%$ to $1.0551\%$. Here, the relative error is defined as $\frac{(t - t^*)}{t}$, which is assumed to follow a normal distribution, as is common practice~\cite{chang2024biathlon}.

}

\section{Experimental Evaluation}
\label{sec:evaluation}

In this section, we present our implementation and compare the end-to-end inference query latency of \CactusDB with baseline systems, followed by an ablation study on the benefits of integrating co-optimization rules. We also evaluate the effectiveness and overheads of our query optimization strategy.

\subsection{Implementation}
\label{sec:implementation}
We implemented our three-level IR and our reusable MCTS query optimizer using Velox~\cite{pedreira2022velox} as a backend. We leveraged Velox's query parser and expression parser to support the top-level and middle-level IR, while implementing all ML functions to support the bottom-level IR on our own. 
The action space of our query optimizer includes all co-optimization rules as discussed in Sec.~\ref{sec:co-optimization-rules}. Our optimizer is applied after rule-based join reordering is executed and before Velox's physical query optimization (e.g., filter reordering and adaptive column prefetching~\cite{pedreira2022velox}) is performed. 
All existing MCTS tree nodes are stored in Faiss~\cite{faiss}, a well-known vector database,  with cosine similarity indexing to accelerate nearest neighbor search.

To generate training data for Model2Vec, we defined a set of model templates—including two-tower recommendation models, Deep Learning Recommendation Model (DLRM)~\cite{naumov2019deep}, autoencoders, singular value decomposition (SVD)~\cite{wall2003singular}, convolutional neural networks, FFNN, embedding models, decision forests, and others (details are provided in Appendix~\ref{sec:app-model-sampling}). From these templates, we randomly sampled $10{,}000$ ML models of varying hyperparameters (e.g., number of neurons, number of layers, number of trees, tree depth, etc.) to construct the training data, with $80\%$ for training and $20\%$ for testing. The trained Model2Vec model is used for embedding the ML expressions to build the feature vectors for the Query2Vec model as described in Sec.~\ref{sec:embedding}. The downstream model inference latency prediction task achieves an average Q-Error~\cite{siddiqui2020cost, zhao2022queryformer} of $1.1$ and a Pearson correlation of $98.02\%$, demonstrating its effectiveness in capturing the complexity of ML models. (Q-Error, defined as $Q(c) = max(\frac{actual(c)}{predicted(c)}, \frac{predicted(c)}{actual(c)})$, is widely used for cost estimation~\cite{siddiqui2020cost, zhao2022queryformer}.)

To train the Query2Vec model, we generated $10{,}000$ training queries using $14$ out of $20$ query templates described in Sec.~\ref{sec:synthetic}. We further split them into $80\%$ training and $20\%$ testing sets to evaluate the model's performance. Our two-model strategy achieved a median Q-Error of $1.19$ and $94\%$ correlation between predicted and actual latency, outperforming the one-model strategy, which achieved a median Q-Error of $1.6$ and $90\%$ correlation. \textcolor{black}{It is important to note that for query optimization, preserving the relative ranking of query plans is more critical than absolute prediction accuracy. The high correlation confirms that our model effectively preserves the relative ranking of candidate query plans. This allows MCTS to reliably explore the search space.}

%




\subsection{Baselines}

We compared the end-to-end query processing latency of \CactusDB to the following baselines: 
\textbf{EvaDB} (v0.3.9), which uses Postgres as backend, provides an AI-centric query optimizer focusing on result caching, query predicate reordering, and parallel query processing \cite{kakkar2023eva, evadb_optimization}. 
\textbf{Apache PySpark w/ UDF} (v3.5.0), which allows model inferences via UDFs~\cite{pyspark, pyspark_model_infer}, with data stored in HDFS. 
\textbf{MADLib} (v2.1.0), which extends Postgres with ML functions~\cite{hellerstein2012madlib}.
\textbf{SystemDS} (v3.2.0), provides a declarative language for data science and ML pipelines, built on Apache Spark~\cite{boehm2019systemds, boehm2016systemml}.
\textbf{PostgresML} (v2.10.0), a Postgres extension for ML~\cite{postgresml}.
{
\textbf{IMBridge}, accelerates in-DB ML inference by caching model contexts and batching predictions efficiently.
}
\textbf{DL-Centric}, where data is stored in Postgres database (v14.15), and then data is transferred into PyTorch (v2.1.2) using ConnectorX (v0.3.2)~\cite{wang2022connectorx} 
for model inferences. Its end-to-end query latency includes both in-database processing time, data transfer time,  and model inference time.


\vspace{3pt}
We also compared the end-to-end latency and the query optimization latency of our reusable MCTS query optimizer with the following optimization algorithms,  all applied after join-reordering and before Velox physical query optimization. 

\noindent
\textbf{Un-optimized}. The query runs without co-optimization.

\noindent
\textbf{Arbitrary Co-optimization.} It scans all co-optimization rules and applies all applicable  rules\cite{zhou2021learned}.

\noindent
\textbf{Heuristics-based.} (1) it will aggressively push down the \texttt{filter}/ \texttt{project} operators; (2) it will aggressively fuse ML operators; (3) Tensor relational transformations are applied only if the model size exceeds a threshold (half of the available memory size).

\noindent
\textbf{Vanilla MCTS.} It builds an MCTS search tree following the naive MCTS search process as described in Sec.~\ref{sec:mcts-vanila}. Both our proposed MCTS optimizer and vanilla optimizer use the learned cost estimation model based on query embeddings, described in Sec.~\ref{sec:embedding}. 

\noindent
\textbf{Reusable.} We evaluated two versions of our reusable MCTS optimizer with the embedding and cost estimation models trained using the one-model and two-model strategies respectively, as described in Sec.~\ref{sec:embedding} and Sec.~\ref{sec:implementation}.

\vspace{3pt}
\noindent
\textcolor{black}{\textbf{Evaluation Environment.} Most experiments were conducted on an AWS r4.2xlarge instance ($8$ CPU cores, $61$ GB memory). The query optimizer evaluation ran on an Ubuntu machine ($48$ CPU cores, $125$ GB memory). GPU experiments used an AWS g4dn.4xlarge instance ($16$ CPU cores, one T4 GPU with $16$ GB memory and $64$ GB host memory).}



\vspace{-5pt}
\subsection{Benchmarks}
\label{sec:exp-benchmark}
%

We used five benchmarks, reflecting realistic industrial query scenarios. These benchmarks nest more than ten diverse AI/ML model architectures with complex SQL queries. Most queries involve up to ten ML functions, $1$-$4$ \texttt{join}/\texttt{crossJoin} operators, $2$-$5$ \texttt{filter}s,  $1$-$7$ \texttt{projection} operators, and up to $4$ \texttt{aggregate}/\texttt{groupBy} operators, ensuring a comprehensive evaluation of co-optimization strategies. See Appendix~\ref{sec:movielens}, ~\ref{appex:tpcxai-qeuries}, and ~\ref{sec:app-query-sampling} for detailed queries.


\subsubsection{Recommendation Queries.} The MovieLens dataset~\cite{harper2015movielens, sarwat2017database} is widely used for evaluating DB-for-ML works~\cite{wooders2023ralf, qin2017scalable, ghorbani2023demonstration, boehm2016systemml, sarwat2017database}. We used  MovieLens-1M dataset, which collected $1$ million ratings from $6{,}000$ users for $4{,}000$ movies and augmented it with movie tag relevance data from MovieLens-32M dataset, which provides $140{,}979$ unique tags and per-movie relevance vectors. We developed three realistic recommendation queries \textcolor{black}{based on existing studies. Q1 is derived from ~\cite{li2022inttower} for pre-ranking systems. It} aggregates user and movie features from ratings, applies an FFNN model and a ``LIKE'' filter to select trending movies within a given movie genre, and cross-joins them with users for scoring via a widely used industrial pre-ranking model, two-tower model~\cite{huang2013learning}.
\textcolor{black}{Q2 and Q3 are adopted from real-world production pipelines described in the multi-stage recommendation technical blog~\cite{nvidia_merlin_multi_stage_recsys} for building intelligent recommender systems.} Q2 applies a movie trending FFNN and a user-interest FFNN to prefilter user-movie pairs, then joins them with movie relevance tag table where 
an AutoEncoder~\cite{hinton2006reducing} is used to transform the high-dimensional movie-tag vector into a low-dimensional dense representation. The output is passed to a DLRM~\cite{naumov2019deep} to estimate a recommendation score.
Q3 uses two FFNN-based filters to identify movies that users may be interested in. For each user-movie pair, a vector similarity search is performed using cosine similarity on the movie's dense representation obtained via the AutoEncoder.

\subsubsection{Retailing-Complex Queries}

\textcolor{black}{To evaluate performance on complex, modern model architectures not currently covered by existing benchmarks, we designed three realistic inference queries for trip classification (Q1), credit card fraud detection (Q2), and product recommendation (Q3) using the TPCx-AI dataset~\cite{brucke2023tpcx} with scale-factors $1$ and $10$. While maintaining the original schema relations, we replaced the simpler default models with production-grade deep learning architectures, including FFNN, XGBoost, and Two-Tower models.} 
%
Q1 joins the \textit{Order} and \textit{Store} tables, applies an ML filter to remove unqualified orders and purchases, and sends the results to an FFNN model for trip classification.
%
Q2 aggregates the \textit{Order} table to generate per-customer features, which is joined with the \textit{Transaction} table and the \textit{Customer} table. The resulting features are sent to an \textit{XGBoost model} and an FFNN model, and return transactions identified as fraud by \textit{both} models. 
Q3 first filters and aggregates the \textit{product\_rating}. The output is joined with the \textit{product} table to embed text attributes as product vectors.
Similar processing is applied to the \textit{customer} table to obtain the customer features. The two are cross-joined and sent 
to a two-tower model that ranks \textit{product-customer} pairs.

\subsubsection{Retailing-Simplified Queries} We found baselines such as MADLib, SystemDS, and PostgresML do not natively support first two benchmarks involving complex AI/ML functions and multiple AI/ML filters. For fair comparisons, we \textcolor{black}{evaluated three simplified queries directly adopted from the official TPCx-AI benchmark with scale-factors $1$ and $10$ for product rating, trip classification, and fraud detection.}
Q1 applies an SVD model~\cite{wall2003singular} to factorize the \textit{product\_rating} matrix for product recommendation. Q2 uses a $50$-tree XGBoost model to classify trips based on aggregated features joined from the \textit{store} and \textit{order} tables. Q3 applies logistic regression for fraud detection by joining the features from \textit{financial\_account} and \textit{financial\_transaction} tables.  

\subsubsection{Other Analytics Queries.} We also evaluated three analytics workloads on real-world datasets - fraud detection on the Credit Card dataset~\cite{kaggle-fraud} (Q1), hotel ranking prediction on Expedia  dataset\textcolor{black}{~\cite{projecthamlet2021}} (Q2), and \text{codeshare} classification on Flights dataset\textcolor{black}{~\cite{projecthamlet2021}} (Q3)\textcolor{black}{, which are widely used for evaluating in-DB ML systems~\cite{park2022end, zhang2024imbridge}.} At initial scale (Scale=1),  the ML inputs have around $289{,}000$ rows and $29$ features for Q1, $79{,}000$ rows and $3{,}000$ features for Q2, and $7{,}000$ rows and $6{,}000$ features for Q3. We further evaluate them at Scale=$10$, which increases the row count by $10\times$.
The Credit Card workload performs a single table scan. Expedia performs a three-way join across \textit{listings}, \textit{hotel}, and \textit{search} tables. Flights performs a four-way join across \textit{routes}, \textit{airlines}, \textit{source}, and \textit{destination airports}. All the workloads apply 4-6 predicate filters, then standardizes numerical and one-hot-encodes categorical columns. The resulting features feed into tree-based classifiers - a single decision tree for Expedia and a $100$-tree ensemble (max depth $9$) for Credit Card and Flights. \textcolor{black}{Additionally, we include the state-of-the-art IMBridge~\cite{zhang2024imbridge, zhang2025mitigating}, as a baseline, which focuses on UDF-level optimization for inference queries. It is excluded from the previous benchmark due to its limited support for complex models used in that settings.}


\subsubsection{A Large-Scale Randomly Generated Inference Query Benchmark}
\label{sec:synthetic}
We designed $20$ inference query templates including \textcolor{black}{$10$ for the MovieLens dataset, and $10$ for TPCx-AI dataset (Scale-$1$). To eliminate selection bias, we derived these templates from community and industry standards. For MovieLens, we crawled the most highly-voted data science notebooks from Kaggle~\cite{movielens2020kaggle} and abstracted their logic.} These templates cover recommendation, rating prediction, collaborative filtering, and rating analysis, using two-tower recommendation models, DLRM, AutoEncoders, SVD, and FFNN. \textcolor{black}{For TPCx-AI, we adopted the queries from its official use cases}, including recommendation, trip classification, fraud and spam detection, customer segmentation, and sales prediction, employing decision forests, SVD, FFNN, logistic regression, and K-Means. Additional details are available in our  Appendix~\ref{sec:app-model-sampling} and \ref{sec:app-query-sampling}. 
Each sampled query varies in model architecture (e.g., different number of layers and neurons for FFNN models), numbers and selectivity of filter predicates.
Following ~\cite{zhao2023comparative}, we divide the query templates into two groups: six randomly chosen from both workloads form the out-of-distribution set used solely for evaluation, while the remaining $14$ generate both training and evaluation queries under different random seeds. We generated $10{,}000$ queries for training the query embedding and cost estimation models, and an additional $1{,}000$ in-distribution and $1{,}000$ out-of-distribution queries to evaluate our optimizer’s generalization capability.



\subsection{Overall Performance Comparison}
\label{sec:overall-perf-comparison}

\begin{table}[h]
\tiny
\centering
\vspace{-3pt}
\caption{\small End-to-End Latency Comparison of Complex Queries}
\label{tab:movielens}

\begin{tabular}{|c|ll|l|l|l|l|l|}
\hline
                            & \multicolumn{2}{l|}{}                                                        & O1                 & O2                 & O3                 & O4                 & Latency                            \\ \hline
\multirow{12}{*}{{\specialcell{Recommendation \\ Queries}}} & \multicolumn{1}{c|}{\multirow{4}{*}{Q1}} & \textbf{\CactusDB} & \cm &                    &                    & \cm & \textbf{0.15} sec \\ \cline{3-8} 
                            & \multicolumn{1}{c|}{}                    & EvaDB                             & \cm &                    &                    &                    & 42.80 sec                          \\ \cline{3-8} 
                            & \multicolumn{1}{c|}{}                    & Apache PySpark w/ UDF             & \cm &                    &                    &                    & 60.28 sec                          \\ \cline{3-8} 
                            & \multicolumn{1}{c|}{}                    & DL-Centric                        &                    &                    &                    & \cm & 48.88 sec                          \\ \cline{2-8} 
                            & \multicolumn{1}{l|}{\multirow{4}{*}{Q2}} & \textbf{\CactusDB} & \cm & \cm & \cm & \cm & \textbf{3.35} sec \\ \cline{3-8} 
                            & \multicolumn{1}{l|}{}                    & EvaDB                             & \cm &                    &                    &                    & Failed (OOM)                       \\ \cline{3-8} 
                            & \multicolumn{1}{l|}{}                    & Apache PySpark w/ UDF             & \cm &                    &                    &                    & Failed (OOM)                       \\ \cline{3-8} 
                            & \multicolumn{1}{l|}{}                    & DL-Centric                        &                    &                    &                    & \cm & 142.96 sec                         \\ \cline{2-8} 
                            & \multicolumn{1}{c|}{\multirow{4}{*}{Q3}} & \textbf{\CactusDB} & \cm &                    & \cm & \cm & \textbf{6.62} sec \\ \cline{3-8} 
                            & \multicolumn{1}{c|}{}                    & EvaDB                             & \cm &                    &                    &                    & Failed (OOM)                       \\ \cline{3-8} 
                            & \multicolumn{1}{c|}{}                    & Apache PySpark w/ UDF             & \cm &                    &                    &                    & Failed (OOM)                       \\ \cline{3-8} 
                            & \multicolumn{1}{c|}{}                    & DL-Centric                        &                    &                    &                    & \cm & 207.12 sec                         \\ \hline
\multirow{12}{*}{\specialcell{Retailing-\\Complex\\ Queries} } & \multicolumn{1}{l|}{\multirow{4}{*}{Q1}} & \textbf{CactusDB}                 & \cm & \cm &                    & \cm & \textbf{72.07 sec}                 \\ \cline{3-8} 
                            & \multicolumn{1}{l|}{}                    & EvaDB                             & \cm &                    &                    &                    & 123.24 sec                         \\ \cline{3-8} 
                            & \multicolumn{1}{l|}{}                    & Apache PySpark w/ UDF                      & \cm &                    &                    &                    & 307.88 sec                         \\ \cline{3-8} 
                            & \multicolumn{1}{l|}{}                    & DL-Centric                        &                    &                    &                    & \cm & 163.97 sec                         \\ \cline{2-8} 
                            & \multicolumn{1}{l|}{\multirow{4}{*}{Q2}} & \textbf{CactusDB}                 & \cm &                    & \cm & \cm & \textbf{44.38 sec}                 \\ \cline{3-8} 
                            & \multicolumn{1}{l|}{}                    & EvaDB                             & \cm &                    &                    &                    & 275.76 sec                         \\ \cline{3-8} 
                            & \multicolumn{1}{l|}{}                    & Apache PySpark w/ UDF                      & \cm &                    &                    &                    & 1320.47 sec                        \\ \cline{3-8} 
                            & \multicolumn{1}{l|}{}                    & DL-Centric                        &                    &                    &                    & \cm & 304.81 sec                         \\ \cline{2-8} 
                            & \multicolumn{1}{l|}{\multirow{4}{*}{Q3}} & \textbf{CactusDB}                 & \cm &                    &                    & \cm & \textbf{3.46 sec}                  \\ \cline{3-8} 
                            & \multicolumn{1}{l|}{}                    & EvaDB                             & \cm &                    &                    &                    & 119.62 sec                         \\ \cline{3-8} 
                            & \multicolumn{1}{l|}{}                    & Apache PySpark w/ UDF                      & \cm &                    &                    &                    & 1525.94 sec                        \\ \cline{3-8} 
                            & \multicolumn{1}{l|}{}                    & DL-Centric                        &                    &                    &                    & \cm & 403.84 sec                         \\ \hline
\end{tabular}                                   
\end{table}

\textbf{Complex Inference Queries.} As shown in Tab.~\ref{tab:movielens}, \CactusDB significantly outperformed the other systems across all six complex queries. \CactusDB achieved $22-401\times$ speedup on recommendation queries, and $1.7-441\times$ speedup on retailing queries, in terms of end-to-end latency compared to the fastest baseline. 

\textcolor{black}{We further analyzed the optimization cost, ranging from $0.02$s to $0.35$s. The specific overhead ratios for the six queries are $13\%$, $3.8\%$, $3.7\%$, $0.5\%$, $0.7\%$, and $0.8\%$, respectively. Notably, across the entire workload, the optimization phase accounts for only approximately $1\%$ of the total cumulative execution time. We further conducted a detailed analysis of the optimizer performance in Sec.~\ref{sec:optimizer-study}.} 

\begin{figure}[ht]
\centering
\includegraphics[width=0.48\textwidth]{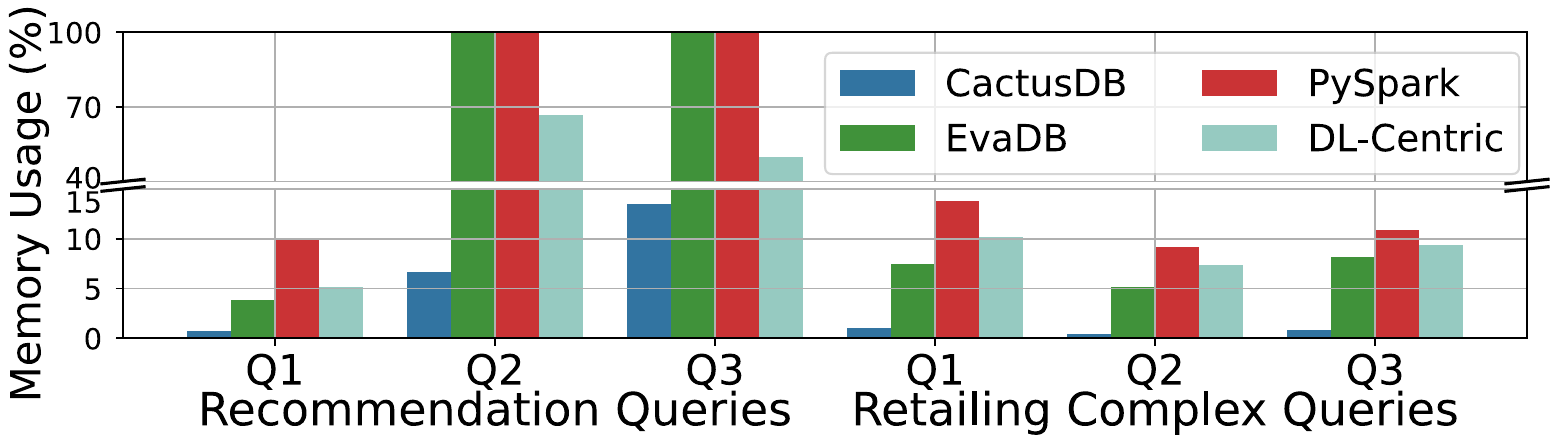}
\caption{\color{black} \small Memory Usage Comparison for Complex Inference Queries \vspace{-3pt}}
\label{fig:table1-memory-comparison}
\end{figure}

\textcolor{black}{Beyond latency, we profiled the peak memory usage to evaluate resource efficiency (Fig.~\ref{fig:table1-memory-comparison}). \CactusDB demonstrated superior memory management, peaking at only $14\%$ usage even for heavy workloads. In contrast, baselines suffered from higher memory usage, which can lead to OOM errors when the model is too large. \CactusDB avoids memory bottlenecks through our query optimizer, which actively selects plans that minimize intermediate data volumes.}

\textbf{(1) Recommendation queries.} For Q1, \CactusDB factorized the two-tower model (R4-1, R1-4) and pushed down the user tower and movie tower through the \texttt{cross-join} (R1-3), leading to a significant reduction in overall latency. For Q1, EvaDB achieved the second-best performance. However, it needs to transfer the data from the database to the Python runtime for inferences, which accounted for $16\%$ of the end-to-end time. For Q2 and Q3, \CactusDB transforms the \texttt{matMul} in a large AutoEncoder model (that encodes the high-dimensional sparse tag matrix into a low-dimensional dense representation) to relational processing (R3-1). This optimization greatly reduced memory footprint. In contrast, EvaDB and the UDF-based PySpark implementation struggled with handling large matrix computations, resulting in out-of-memory (OOM) errors. DL-Centric implementation suffers from cross-system feature transfer overhead, which accounts for $37\%$ of end-to-end time. 

\textbf{(2) Retailing-Complex queries.} \CactusDB significantly outperforms all baselines across the three queries, achieving the highest speedup on Q3 (product recommendation). The gain is mainly driven by splitting user-product feature computations(R4-1) and pushdown (R1-3). For trip type classification (Q1) and fraudulent transaction detection (Q2), performance gain stems from factorizing \texttt{matMul} (R2-1) and the transformation of XGBoost to \texttt{crossJoin} and \texttt{aggregate} (R3-2), respectively. Additional improvements come from reordering and pushing down AI/ML filters (R1-1 and R1-2). In contrast, EvaDB and DL-Centric incur high latency from database-Python transfer and costly operators, such as matrix multiplication and XGBoost, while UDF-based PySpark performs worst due to missing UDF optimizations.

\begin{figure}[ht]
\centering
\color{black}
\includegraphics[width=0.48\textwidth]{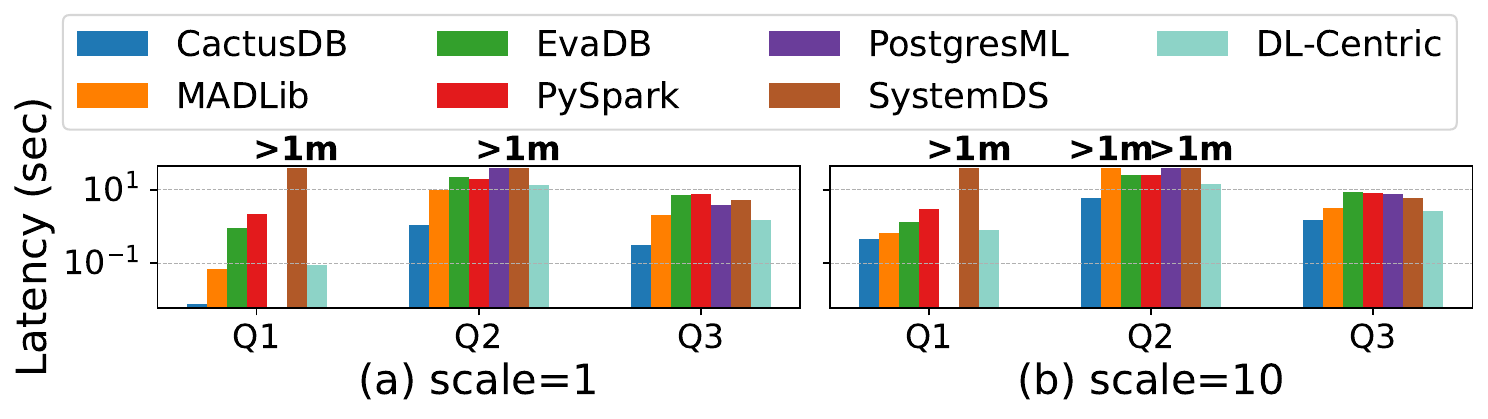}
\caption{\color{black} \small Latency Comparison for Retailing-Simplified Queries (Log Scale). "\textgreater1m" indicates latency greater than 1 minute. \vspace{-3pt}}
\label{fig:result-tpcxai-simple}
\end{figure}

\noindent
\textbf{ Retailing-Simplified Queries.} Fig.~\ref{fig:result-tpcxai-simple} shows that {{\CactusDB consistently achieved the best performance across all three queries at both scale factors, yielding $1.7-20.8\times$ speedups over the fastest baseline}}. MADLib outperforms well for Q1 and Q2 at scale factor $1$, but it internally relies on Python runtime (e.g., scikit-learn's SVD and XGBoost) introduces extra overhead. 
In Q2 with scale factor $10$, MADLib introduces additional recursive function calls during XGBoost inference as the data size increases, leading to significant performance degradation. In Q3, its native logistic regression is less efficient than scikit-learn's highly optimized implementation. For PostgresML, its batch prediction API is significantly faster than its regular prediction API but implemented as an aggregation operator, preventing simultaneous access to non-group-by attributes and causing inefficiency in Q2. It also lacks SVD model support for Q1.
SystemDS exhibits slower inference due to less optimized XGBoost and SVD implementations, while DL-Centric suffers from data transfer overhead despite leveraging high-performance ConnectorX library~\cite{wang2022connectorx}.

\begin{figure}[ht]
\centering
\color{black}

\includegraphics[width=0.48\textwidth]{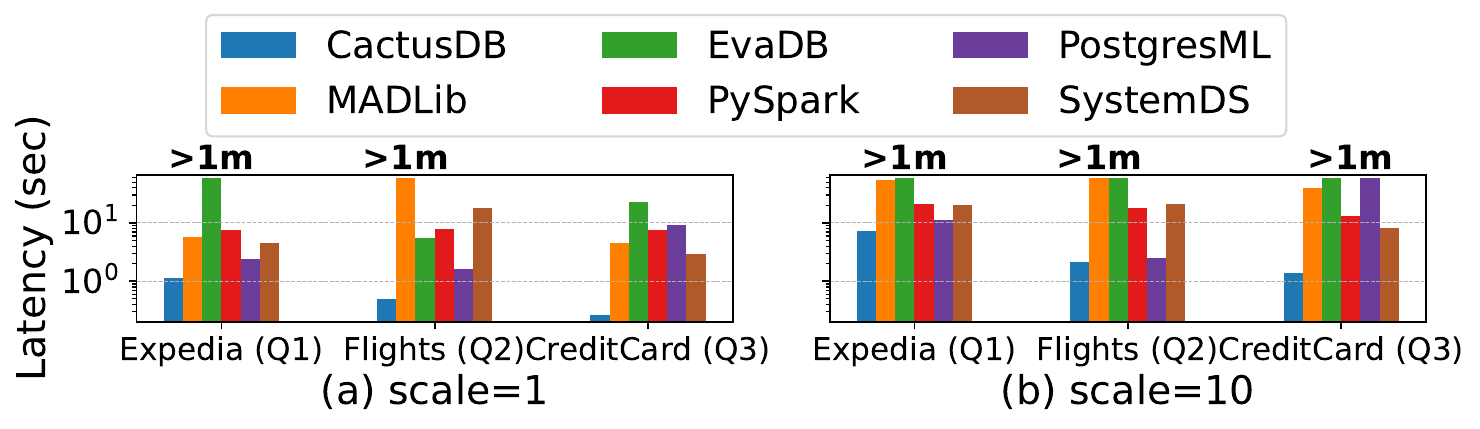}

\caption{\color{black} \small Latency Comparison for Three Analytics Queries (Log Scale). "\textgreater1m" indicates latency greater than 1 minute.}
\vspace{-5pt}
\label{fig:analytics-queries}
\end{figure}

\vspace{-3pt}
\noindent
\textbf{Other Analytics Queries.} 
As shown in Fig.~\ref{fig:analytics-queries}, {{\CactusDB outperforms the best of other baselines by $2-2.5\times$ at scale-$1$ and $1.2-5.7\times$ at scale-$10$.}} 
PostgresML is slightly slower ($1.5\times$ and $1.2\times$ slower) than \CactusDB on the first two queries on scale-10 data, but struggles with the larger Q3 workload that contains many rows, on which it is $66\times$ slower than \CactusDB. \textcolor{black}{IMBridge benefits from batch processing and context caching but suffers from data transfer overhead between the database query engine and the model inference backend when data size increased.}
Although Pyspark and SystemDS benefit from their parallel execution and exhibit more stable runtimes, \CactusDB outperformed PySpark by $1.6-9.2\times$ and SystemDS by $1.6-9.5\times$ on scale-10 data. 
EvaDB and MADLib underperform other baselines for these workloads. 

\noindent
\textbf{Ablation Study.}
To understand the benefits of integrating all co-optimization rules, we compare the performance of applying all optimization techniques versus only one category 
on the recommendation and retailing-complex queries.
As shown in Tab.~\ref{tab:exp-ablation}, {{integrating all co-optimization rules yields substantially higher speedups than using only one category
}}. For example, Recommendation-Q1 achieves $676.1\times$ speedup over the default plan by pushing down filters and factorizing model computation to avoid redundant inference.


\begin{table}[h]
\tiny
\centering
\caption{\small Ablation Study}
\label{tab:exp-ablation}
\begin{tabular}{|c|c|c|c|c|c|c|c|}
\hline
    Workloads                       & Queries & Un-optimized & O1    & O2    & O3   & O4    & Combined \\ \hline
\multirow{3}{*}{Recommendation} & Q1      & 1.0X         & 45.1X & NA    & NA   & 10.9X & 676.1X   \\ \cline{2-8} 
                           & Q2      & 1.0X         & 3.8X  & 2.9X  & 1.8X & 2.0X  & 18.5X     \\ \cline{2-8} 
                           & Q3      & 1.0X         & 3.5X  & NA    & 2.1X & 1.5X  & 12.4X     \\ \hline
\multirow{3}{*}{Retailing} & Q1      & 1.0X         & 1.3X & 1.5X & NA   & 1.2X  & 1.9X     \\ \cline{2-8} 
                           & Q2      & 1.0X         & 1.7X & 2.0X & 1.5X & 1.6X & 4.3X    \\ \cline{2-8} 
                           & Q3      & 1.0X         & 2.9X & 15.8X & NA   & 2.6X  & 28.3X    \\ \hline
\end{tabular}
\vspace{-8pt}
\end{table}


\begin{figure}[ht]
\centering

\includegraphics[width=0.52\textwidth]{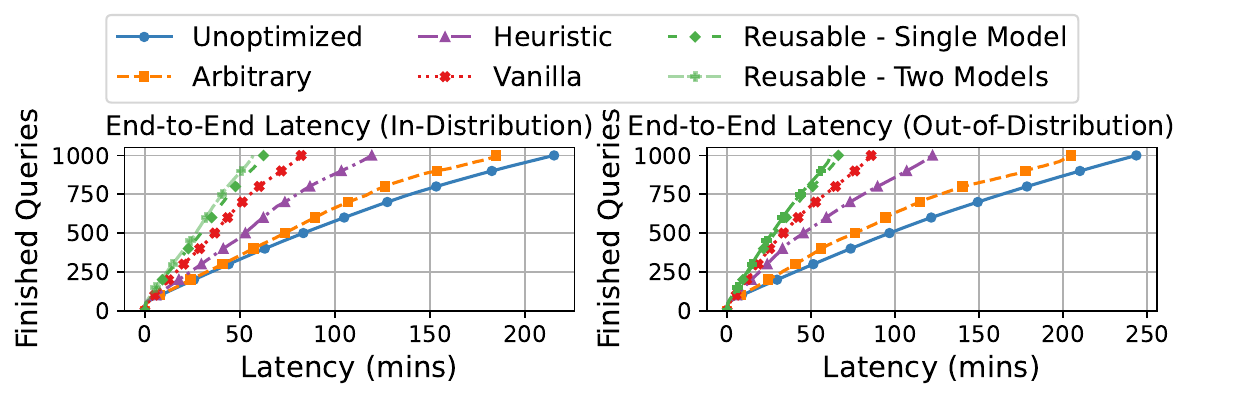}

\hspace{0.5pt}
\vspace{-8pt}
\caption{\color{black} \small Query Optimizer Comparison on End-to-End Latency}
\label{fig:optimizer-e2e}
\end{figure}

\begin{figure}[ht]
\centering
\vspace{-5pt} 

\includegraphics[width=0.51\textwidth]{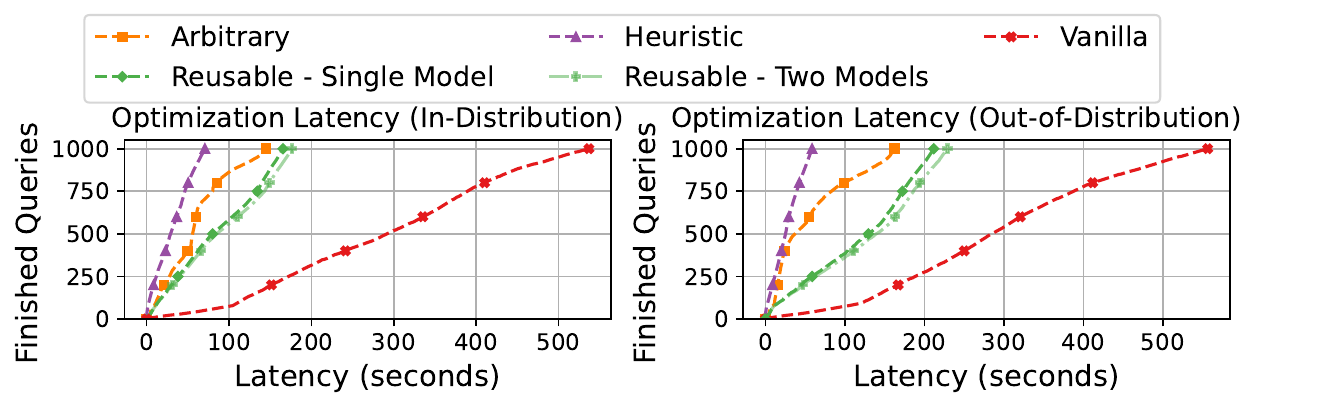}

\hspace{0.5pt}
\vspace{-8pt}
\caption{\color{black} \small Query Optimizer Comparison on Optimization Latency}
\label{fig:optimizer-opt}
\end{figure}





\subsection{Evaluation of Reusable MCTS Optimizer}
\label{sec:optimizer-study}

{
\color{black}
\noindent
In this section, we compare our reusable MCTS inference query optimizer with alternative optimization strategies in terms of end-to-end latency and optimization overhead on $1{,}000$ in-distribution (ID) queries and $1{,}000$ out-of-distribution (OOD) queries. Results are reported as elapsed latency versus the number of finished queries in Fig.~\ref{fig:optimizer-e2e} and Fig.~\ref{fig:optimizer-opt}.
The results revealed that {the reusable MCTS effectively balances optimization overheads and execution latency by reusing MCTS search trees across queries.} 

For ID queries, the reusable MCTS achieves an $89\%$ state collision rate (i.e., the ratio of queries matching existing MCTS states)
greatly reducing optimization time compared to vanilla MCTS. While, vanilla MCTS produces high-quality query plans, it incurs substantial optimization overhead due to the need of rebuilding a search tree from scratch for each query, becoming a bottleneck as shown in Fig.~\ref{fig:optimizer-opt}. The heuristic optimizer performs better than the arbitrary baseline but often produces suboptimal plans due to its fixed rule order.

Our reusable MCTS also generalizes well to OOD queries. Initially, OOD queries require building new MCTS trees due to the absence of reusable states, resulting in higher optimization latency, but optimization efficiency improves as more reusable states are found. 
On average, the state collision rate reaches $72\%$, indicating strong generalization.
As shown in Fig.~\ref{fig:optimizer-e2e} and Fig.~\ref{fig:optimizer-opt}, our two-model strategy introduces minor optimization overhead
—$6\%$ for ID and $7\%$ for OOD queries.
However, the improved query plan latency estimation leads to better optimization results, achieving $10\%$ and $7\%$ improvements in end-to-end execution latency, respectively.

}

{
\color{black}
\noindent
\textbf{Storage Overhead Analysis} 
The Model2Vec and Query2Vec models occupy $8$MB and $23$MB, respectively. We incrementally maintain MCTS search trees, where each node requiring about $1.6$KB of storage. 
Thanks to MCTS tree reusability, the overall storage cost remains low. For instance, optimizing $1{,}000$ queries requires less than $2$MB in total.
}

\noindent
{\color{black}
\section{Related Work}
\label{sec:related-works}
Prior in-database AI/ML systems and unified intermediate representations (IRs) support only
subsets of SQL-ML co-optimization techniques (O1-O4).
UDF-centric systems such as EvaDB~\cite{kakkar2023eva}, PostgresML~\cite{postgresml},
MADLib~\cite{hellerstein2012madlib}, and PySpark~\cite{singh2022manage} primarily enable
O1-style relational rewrites while treating AI/ML logic as opaque, which precludes deeper
co-optimization.
Factorized ML systems~\cite{factorize-norm-data, factorize-lmfao, factorize-joinboost,
factorize-la, li2019enabling, schleich2016learning} support O2-style optimization for
predefined models or linear algebra operators, but focus mainly on training and lack
integration with general inference pipelines.
Other systems and IRs, including SimSQL~\cite{luo2018scalable, jankov12declarative},
SystemML/SystemDS~\cite{boehm2016systemml, boehm2019systemds},
MASQ~\cite{paganelli2023pushing, del2021transforming},
Raven~\cite{park2022end},
SDQL~\cite{shaikhha2022functional},
Weld~\cite{palkar2017weld, palkar2018evaluating},
Lara~\cite{kunft2019intermediate},
LingoDB~\cite{jungmair2022designing, jungmair2023declarative},
and Tensor Relational Algebra~\cite{DBLP:journals/pvldb/YuanJZTBJ21},
combine relational, linear algebra, and ML operators to varying degrees, yet none of them support
flexible decomposition of inference pipelines with end-to-end optimization across O1-O4.

Existing inference optimizers rely largely on rule-based, cost-based, or learned strategies
(e.g., Spark Catalyst~\cite{pirahesh1992extensible, armbrust2015spark},
Cascades-style CBOs~\cite{graefe1995cascades, calcite, postgresqldb},
and learned optimizers~\cite{gaussml, zhou2021learned, marcus2019neo, marcus2021bao}),
which struggle with arbitrary AI/ML workloads and ad-hoc inference queries.
In contrast, \CactusDB is a unified in-database inference system that supports relational,
linear-algebra, and ML operators within a three-level IR, decomposes inference pipelines into
fine-grained sub-computations, and employs a learning-assisted optimizer to efficiently explore
a large co-optimization search space, enabling systematic composition of O1-O4 techniques.
}

\section{Conclusion}
\label{sec:conclusions}
In this paper, we present \CactusDB, a system that enables efficient execution and automatic optimization of SQL-ML queries. 
\CactusDB is the first database system that supports applying diverse co-optimization techniques to different sub-computations of one inference query by introducing a flexible and expressive three-level IR.
To efficiently explore the enlarged search space resulting from the integration of co-optimization techniques, we propose a novel, reusable MCTS-based optimizer co-designed with our three-level IR. Unlike existing MCTS-based query optimizers, ours represents query plans as embedding vectors (states) and co-optimization techniques as configurable actions, enabling MCTS state sharing across queries. 
%
%
\CactusDB achieved $1.7\times$ to $441\times$ speedup over the fastest baseline systems across $11$ inference workloads. Evaluation on $2{,}000$ inference queries further demonstrates that our reusable MCTS optimizer achieved better effectiveness-efficiency trade-offs than alternative optimizers. 

{\color{black}
\noindent
\textbf{Limitations and Future Works}
First, \CactusDB currently lacks a user-friendly front-end for converting existing Python-based data science pipelines into our three-level IR forms. As a result, integrating legacy workflows requires manual effort and a learning curve.
Second, similar to Raven~\cite{park2022end}, the \CactusDB query optimizer relies on a predefined set of machine learning functions and rewrite rules to support various co-optimization techniques. Introducing new co-optimization strategies that involve novel AI/ML functions requires additional engineering effort to implement the corresponding functions and explicitly inform the optimizer of the associated rewrite actions.
Third, the current implementation represents complex models, such as large language models (LLMs), as black boxes, due to the challenge of converting these models to compact computational graphs. Consequently, \CactusDB can only apply a subset of co-optimization techniques to these models.

Future work includes (1) developing user-friendly toolchains to facilitate the import of legacy code and the extension of the query optimizer to new ML functions and rewrite rules, and (2) representing LLMs as white-box models to enable more effective co-optimization techniques.
}

\section{AI-Generated Content Acknowledgement}

No generative AI tools were used to generate the content in this submission, including text, figures, tables, and code.

\section{Acknowledgments}
We sincerely thank Saif Masood and Qi Lin for their contributions to the development of ML functions and rewrite actions at an early stage of this project. We also thank the anonymous reviewers from VLDB'25, SIGMOD'26, and ICDE'26 for their constructive and insightful feedback. This work was supported by the National Science Foundation (NSF) CAREER Award (No.~2144923) and an Amazon Research Award.

\bibliographystyle{IEEEtran}
\bibliography{refs}

\clearpage

\newpage

\appendix

{\color{black}{
\subsection{An Extensive List of SQL-ML Co-Optimization Techniques}
\label{sec:extensive-rules}

We summarize (non-approximate) SQL-ML co-optimization techniques  into four main categories, O1-O4. We also highlight the disparate data abstraction (focusing on model parameter representation) and computation abstraction across these categories.

\vspace{3pt}
\noindent
 \textbf{O1. Nested Relational Algebra Optimization}~\cite{evadb_optimization, gaussml, postgresml, hellerstein2012madlib}.  The \underline{computation abstraction} consists of relational operators such as \texttt{filter}, \texttt{project}, \texttt{join}, \texttt{crossJoin}, and \texttt{aggregate}. Each operator can be customized by an expression consisting of expression operators, such as =, >, <, $\wedge$, \texttt{CALLFUNC} (i.e., invoking an arbitrary opaque function for feature processing, AI/ML model inferences, etc.).  In the \underline{data abstraction}, each weight tensor used by the AI/ML models is encapsulated in the functions, opaque to the query optimizer.
\underline{Example co-optimization techniques} in O1 include:

\begin{shaded}{
\smaller

\noindent
(\textbf{Notations:} $A$ and $B$ are two attributes of the table $R$, of \texttt{vector} type and \texttt{float} type respectively, and $k$ is a constant.)

\noindent
\textbf{R1-1} \textbf{\texttt{filter} Reorder.} It reorders two \texttt{filter}s that are customized by user-defined expressions such as $dnnFraudDetect(R.A) = True$ and $R.B > k$. so that the one with smaller selectivity is executed first to reduce the amount of data input to the other.

\noindent
\textbf{R1-2} \textbf{\texttt{filter} Expression Merge and Factorization.} Two \texttt{filters}  with different boolean expressions on the same dataset, can be merged into one filter with combined expressions in which redundant computations can be removed, e.g.,  $f(R.A) = True \wedge R.B > k$ and $R.B < k-1$ (a naive example) can be merged into $false$. Factorization is the reverse process of merging.

\noindent
\textbf{R1-3} \textbf{\texttt{filter} Pushdown/Pullup.} A \texttt{filter} can be pushed down through a \texttt{join} to avoid redundant evaluation of the \texttt{filter}'s expression over repeated records caused by the \texttt{join}, if the \texttt{join}'s output has a high cardinality. Otherwise, the \texttt{filter} should be pulled up to avoid evaluating expressions over records that do not have a match through the \texttt{join}.

\noindent
\textbf{R1-4} \textbf{\texttt{project} Merge and Factorization.} Similar to R1-2, two consecutive \texttt{project}s customized by expressions can be merged into one \texttt{project} with optimized expressions. Factorization is the reverse process of merging.

\noindent
\textbf{R1-5} \textbf{\texttt{project} Pushdown/Pullup.} Similar to R1-3, \texttt{project} with expressions can be pushed down/pulled up to avoid redundant/unnecessary computations.
}
\end{shaded}


\vspace{3pt}
\noindent
 \textbf{O2. Factorized Inference} (derived from factorized ML~\cite{factorize-la, li2019enabling, factorize-lmfao}). The \underline{computation abstraction} consists of relational operators and a limited set of factorizable ML functions, such as inner product of vector and weight matrix. 
For \underline{data abstraction}, parameters of a factorizable ML function are stored as a factorizable object, visible to the query optimizer.
\underline{Example co-optimization techniques} include:

\begin{shaded}{
\smaller
\noindent
(\textbf{Notations:} $S(Y, \boldsymbol{X}_S, FK)$ and $R(\underline{RID}, \boldsymbol{X}_R)$ are two tables. $\boldsymbol{X}_S$ and $\boldsymbol{X}_R$ are the feature vectors. $RID$ is the key attribute of $R$, and $FK$ is a foreign key of $S$ referencing $RID$. $Y$ is the union of rest columns in $S$. After joining $S$ and $R$ on $FK=RID$, we can obtain an output table $T(Y, \boldsymbol{X})$, where $\boldsymbol{X}\equiv [\boldsymbol{X}_S, \boldsymbol{X}_R]$ is the concatenation of the feature vectors.)

\noindent
 \textbf{R2-1} \textbf{Linear Algebra Operator Factorization}~\cite{factorize-la, li2019enabling, factorize-lmfao}. Taking inner product for example,  for each feature vector $x \in T.\boldsymbol{X}$, we perform an inner product $w^Tx$ as part of the inference computation, where $w$ is a weight matrix. Since $\boldsymbol{T}$ has redundant tuples, redundant inner product computation will occur, which can be avoided by factorizing the inner products over $x \in \boldsymbol{X}$ into inner products over the feature vectors $x_R \in \boldsymbol{X}_S$ and $x_R \in \boldsymbol{X}_R$, leveraging $w^Tx = w^T_Sx_S + w^T_Rx_R$, wherein $w_S/w_R$ is the projection of $w$ to the features from $S/R$. Pushing $w^T_Sx_S$ down to table $S$ and pushing  $w^T_Rx_R$ down to table $R$ will avoid redundant computations caused by the join.
      
\noindent
\textbf{R2-2 Factorization of Decision Tree Inference}~\cite{lucchese2015quickscorer, lucchese2017quickscorer}. 
        A $k$-depth decision tree $f$ contains $2^{k-1}$ leaf nodes and $2^{k-1}-1$ intermediate nodes. Each intermediate node is associated with a feature in the input $x \in T.\boldsymbol{X}$, and outputs True/False based on the feature's value. Intermediate nodes are divided and pushed down to $R$ or $S$ based on whether their features are from $R$ or $S$. The detailed factorized inference process is based on the QuickScorer algorithm~\cite{lucchese2015quickscorer, lucchese2017quickscorer} that encodes each intermediate node into a $2^{k-1}$-length bit vector and converts the decision tree inference result into the bitwise AND of the bit vectors of all False nodes (i.e., intermediate tree nodes that return False on the input feature). The bitwise AND computations can be performed on $R$ and $S$ respectively, and then aggregated. 

\noindent
\textbf{R2-3} \textbf{Factorization of Distance Computation}~\cite{gaussml}. If each feature vector $x \in \boldsymbol{X}$ needs to compute a Euclidean distance with a vector of same length called $y$, i.e., $dist(x, y)$, we can avoid redundant computation on repeated tuples in $T$ by factorizing the distance computation into $dist(x_S, y_S)$ and $dist(x_R, y_R)$, wherein $y_S/y_R$ is the projection of $y$ to the features from $S/R$, and pushing down them to $S$ and $R$ respectively, leveraging $dist(x, y)=\sqrt{dist^2(x_S, y_S)+dist^2(x_R, y_R)}$. }
        %
\end{shaded}


\vspace{3pt}
\noindent
 \textbf{O3. Tensor-Relational Transformation.}
 The \underline{computation abstr-} \underline{action} consists of a limited set of tensor-relational operators~\cite{DBLP:journals/pvldb/YuanJZTBJ21}, such as \texttt{aggregate}, \texttt{join}, \texttt{crossJoin}, \texttt{filter}, \texttt{project}, \texttt{rekey}, \texttt{tile}, \texttt{concat}, all processing relations. 
In the \underline{data abstraction}, the model parameters must be transformed into relations (as discussed in R3-1 to R3-3). 
\underline{Example co-optimization techniques} include:

\begin{shaded}
\smaller
\noindent
\noindent
(\textbf{Notations:} similar to the notations in O2, $\boldsymbol{T}(Y, \boldsymbol{X})$ is a relation, where $\boldsymbol{X}$ is a column of feature vectors.)

\noindent
\textbf{R3-1} \textbf{Linear algebra operators transformed to relational operators}~\cite{DBLP:journals/pvldb/YuanJZTBJ21, luo2018scalable, boehm2016systemml, DBLP:journals/pvldb/ZhouCDMYZZ22}. 
To enable the transformation, a matrix $w$ could be transformed into a relation $R(\underline{colId}, \boldsymbol{tile})$, where each tuple represents a vertically partitioned matrix tile indexed by $colId$, with its data flattened in a 1-D array in the $\boldsymbol{tile}$ attribute.
Then taking \texttt{matMul} between feature vectors in $T.X$ and $w$ as an example, it can be converted into a \texttt{crossJoin} followed by \texttt{project}. The \texttt{crossJoin} pairs each feature vector in $T$ with each tile in $R$. Then each joined pair of $x \in T.\boldsymbol{X}$ and $t \in R.\boldsymbol{tile}$ is applied with a \texttt{project} that computes $x \times t$. An output block has its row index being the feature vector's row index and its column index being the tile's column index. 
As described in Tensor Relational Algebra (TRA)~\cite{DBLP:journals/pvldb/YuanJZTBJ21}, other linear algebra computations can also be represented in relational algebra. 


\vspace{2pt}
\noindent
\textbf{R3-2} \textbf{Decision Forest, e.g., \texttt{XGBoost} and \texttt{lightGBM}, to \texttt{crossJoin}, \texttt{project}, and \texttt{aggregate}}~\cite{guan2023comparison}. A decision forest model can be represented as a relation $DF(TreeId: INT, TreeObj: TreeType)$, where each tuple represents a decision tree indexed by $TreeId$, and stored as a tree object in the $TreeObj$ column. Applying the model to each feature vector $x \in T.\boldsymbol{X}$, could be transformed into relational processing by first performing a \texttt{crossJoin} of $T$ and $DF$. Then a \texttt{project} is applied to each joined pair of $x \in T.\boldsymbol{X}$ and $t \in DF$ to run $t.predict(x)$. Furthermore, the prediction results are aggregated to obtain the final prediction.

\vspace{2pt}
\noindent
\textbf{R3-3} \textbf{\texttt{distances\_to\_centroids} to \texttt{crossJoin} and \texttt{project}.} Given a set of clusters represented in a relation $R(ClusterId, \boldsymbol{C})$, where each tuple represents a cluster indexed by $ClusterId$ with centroid $c$ in the $\boldsymbol{C}$ column. Computing the distance of each feature vector $x \in T.\boldsymbol{X}$ with each centroid  $c \in R.\boldsymbol{C}$  can be represented as a \texttt{cross\_join} of $T$ and $R$, followed by a \texttt{project} that computes the distance between $c \in R.\boldsymbol{C}$ and $x \in T.\boldsymbol{X}$.

\end{shaded}


The O3 transformations scale to AI/ML models that involve large-scale parameters, by blocking the parameters so that each time only a few blocks need to be loaded into the memory (via buffer pool management). However, such transformation increases overheads for AI/ML models with small parameters~\cite{zhou2024serving}.

\vspace{3pt}
\noindent
 \textbf{O4. Data-Driven Optimization of AI/ML Models~\cite{park2022end}.}
 The \underline{computation abstraction} of O4 consists of ML functions, each corresponding to a node in a computational graph. 
The \underline{data abstraction} is tensor objects, corresponding to each edge in a computational graph. 
\underline{Example optimization techniques} in O4 include:

\begin{shaded}
\smaller
\noindent
\textbf{R4-1} \textbf{AI/ML Operator Fusion and Factorization.} Multiple consecutive AI/ML operators can be fused into a monolithic operator to reduce function call overhead. Factorization is the reverse of the fusing process.
          

\vspace{2pt}
\noindent
\textbf{R4-2} \textbf{AI/ML Library Replacement.} An AI/ML operator's implementation can be replaced by different libraries. For example, for \texttt{matMul}, there are different implementations for CPU, GPU, sparse matrices, and dense matrices. 

\vspace{2pt}
\noindent
\textbf{R4-3} \textbf{AI/ML Algorithm Replacement.} For example, a \texttt{conv2D} operator can be converted into \texttt{matMul} through spatial rewriting of the input images and the filters ~\cite{vasudevan2017parallel, Conv-spatial-rewrite, spatial-rewrite}.
        The detailed process is explained in several references~\cite{vasudevan2017parallel, Conv-spatial-rewrite, spatial-rewrite}. 


\vspace{2pt}
\noindent
\textbf{R4-4} \textbf{Constant Folding.} If an operator $f$ always has an input $c$ (determined through data profiling in the database), it can be replaced with the precomputed constant 
$f(c)$, eliminating the need for runtime evaluation. 

\end{shaded}

In this paper, we argue that it is important to support all optimization techniques from O1-O4 in one database system. First, it will enable different sub-computations to be applied with different data abstractions and co-optimization techniques, maximizing flexibility. Second, we found combining optimization techniques from different categories will unlock new optimization opportunities. For example, applying R4-1 to factorize a high-level ML function into atomic ML functions such as \texttt{matMul} will create new opportunities for applying R2-1 and R1-3 to factorize and push-down the \texttt{matMul} or applying R3-1 to convert the \texttt{matMul} into relational operators.  

\begin{figure}[t] 
	\setlength\tabcolsep{1pt} 
	\centering	
    \color{black}
	\resizebox{1\columnwidth}{!}{ \small 
	\begin{tabular}
		{|c|c|p{50mm}|c|c|c|c|c|c|c|c|c|c|c|c|c|c|c|} 
		
		\hline
		
		\hline
		
		\multicolumn{3}{|c|}{\cellcolor{lightgray}{\multirow[c]{1}{*}{\cellcolor{lightgray}{\rotatebox[origin=lb]{0}{{\textbf{Legend}}}}}}} & 
		
		\multicolumn{5}{c|}{{\multirow[c]{1}{*}{\rotatebox[origin=lb]{0}{{\scriptsize O1}}} }} &
		\multicolumn{3}{c|}{{\multirow[c]{1}{*}{\rotatebox[origin=lb]{0}{{\scriptsize O2}}}}} &
		\multicolumn{3}{c|}{{\multirow[c]{1}{*}{\rotatebox[origin=lb]{0}{{\scriptsize O3}}}}} &
		\multicolumn{4}{c|}{{\multirow[c]{1}{*}{\rotatebox[origin=lb]{0}{{\scriptsize O4}}}}} \\
  
		\hline
		
		\hline
        
		\multicolumn{2}{|r|}{\cellcolor{lightgray}{\small  	}} & 			\multicolumn{1}{|l|}{\cellcolor{lightgray}\small Not applicable}		&&&&&&&&&&&&&&&\\
		\multicolumn{2}{|r|}{\cellcolor{lightgray}{\small !	}} & 			\multicolumn{1}{|l|}{\cellcolor{lightgray}\small Partially support } &&&&&&&&&&&&&&&\\
        \multicolumn{2}{|r|}{\cellcolor{lightgray}{\small $\oplus$	}} & 			\multicolumn{1}{|l|}{\cellcolor{lightgray}\small Supported by external extension } &&&&&&&&&&&&&&&\\
        \multicolumn{2}{|r|}{\cellcolor{lightgray}{\small \cm	}} & 			\multicolumn{1}{|l|}{\cellcolor{lightgray}\small Fully supported} & 
		\multirow[t]{4}{*}{\begin{sideways}{{\small R1-1}}\end{sideways}} &
  		\multirow[t]{4}{*}{\begin{sideways}{{\small R1-2}}\end{sideways}} &
            \multirow[t]{4}{*}{\begin{sideways}{{\small R1-3}}\end{sideways}} &
		\multirow[t]{4}{*}{\begin{sideways}{{\small R1-4}}\end{sideways}} &
            \multirow[t]{4}{*}{\begin{sideways}{{\small R1-5}}\end{sideways}} &
		\multirow[t]{4}{*}{\begin{sideways}{{\small R2-1}}\end{sideways}}&
		\multirow[t]{4}{*}{\begin{sideways}{{\small R2-2}}\end{sideways}} &
		\multirow[t]{4}{*}{\begin{sideways}{{\small R2-3}}\end{sideways}} &
		\multirow[t]{4}{*}{\begin{sideways}{{\small R3-1}}\end{sideways}} &
		\multirow[t]{4}{*}{\begin{sideways}{{\small R3-2}}\end{sideways}} &
		\multirow[t]{4}{*}{\begin{sideways}{{\small R3-3}}\end{sideways}} &
		\multirow[t]{4}{*}{\begin{sideways}{{\small R4-1}}\end{sideways}} &
            \multirow[t]{4}{*}{\begin{sideways}{{\small R4-2}}\end{sideways}} &
		\multirow[t]{4}{*}{\begin{sideways}{{\small R4-3}}\end{sideways}} &
		\multirow[t]{4}{*}{\begin{sideways}{{\small R4-4 }}\end{sideways}} \\
        
		\hline 
        
		\hline
        
		\multirow{4}{*}{\begin{sideways}{UDF-centric}\end{sideways}} & 
		 \multicolumn{2}{|l|}{\textbf{CactusDB}}									&\cm&\cm&\cm&\cm&\cm&\cm &\cm&\cm&\cm	&\cm&\cm&\cm&\cm&\cm&\cm\\
		\cline{\sln-18}& \multicolumn{2}{|l|}{MADLib~\cite{hellerstein2012madlib}}									& ! & !& !& !& !&&	\multicolumn{1}{c|}{} &&	&	&		& & !&		&\\
		\cline{\sln-18}& \multicolumn{2}{|l|}{EvaDB~\cite{kakkar2023eva}} 										&\cm&\cm& ! &\cm& &&	\multicolumn{1}{c|}{} &&		&&		& $\oplus$ & $\oplus$ &		&$\oplus$\\
      		\cline{\sln-18}& \multicolumn{2}{|l|}{PySpark-UDF~\cite{singh2022manage}} 										&!&!&!&!&&&	\multicolumn{1}{c|}{} &&!		&!&!		&$\oplus$ &$\oplus$&		&$\oplus$\\	
            
		\hline					
        
		\hline

  		\multirow{5}{*}{\begin{sideways}{Factorized}\end{sideways}} & 
		\multicolumn{2}{|l|}{Factorized Learning~\cite{factorize-norm-data}} 																&&&&&\cm& \cm &	\multicolumn{1}{c|}{} &&		&&		& &&		&\\
		\cline{\sln-18}& \multicolumn{2}{|l|}{F/SQL~\cite{schleich2016learning}} 																	&&&&&\cm& \cm &	\multicolumn{1}{c|}{} &&		&&		& &&		&\\
		\cline{\sln-18}& \multicolumn{2}{|l|}{LMFAO~\cite{factorize-lmfao}} 																&&&&&\cm& ! &	\multicolumn{1}{c|}{\cm} &&		&&		& &&		&\\
		\cline{\sln-18}& \multicolumn{2}{|l|}{JoinBoost~\cite{factorize-joinboost}} 																&&&&&\cm&& \multicolumn{1}{c|}{\cm} && 	&&		& &&		&\\
		\cline{\sln-18}& \multicolumn{2}{|l|}{Morpheus~\cite{factorize-la, li2019enabling}} 																&&&&&\cm& \cm &	\multicolumn{1}{c|}{} & \cm &		&&		& &&		&\\
        
             \hline	
             
		\hline
		
		\multirow{4}{*}{\begin{sideways}{RA-centric}\end{sideways}} & 
						   \multicolumn{2}{|l|}{\cellcolor{lightgray}SimSQL~\cite{luo2018scalable}} 																&&&&&&&	\multicolumn{1}{c|}{} &&	\cm	&&\cm		& &&		&\\
		\cline{\sln-18}& \multicolumn{2}{|l|}{\cellcolor{lightgray}SystemML/SystemDS~\cite{boehm2016systemml, boehm2019systemds}}																&&&&&& &	\multicolumn{1}{c|}{} &&\cm		&!&\cm		&! &!&		&\\
  
      \cline{\sln-18}&\multicolumn{2}{|l|}{\cellcolor{lightgray}TRA~\cite{DBLP:journals/pvldb/YuanJZTBJ21}}																&&&&&& &	\multicolumn{1}{c|}{} &&\cm		&&\cm	& &&		&\\
      \cline{\sln-18}& \multicolumn{2}{|l|}{MASQ~\cite{paganelli2023pushing, del2021transforming}}																&&&\cm&&\cm&&	\multicolumn{1}{c|}{} &&	&\cm&		&!  &!&		&\\
      
		\hline	
        
		\hline
        
  		\multirow{3}{*}{\begin{sideways}{SQLext}\end{sideways}} & 
						   \multicolumn{2}{|l|}{GaussML~\cite{gaussml}} 																&\cm&\cm&\cm&\cm&\cm&&	\multicolumn{1}{c|}{} &\cm&		&&		& &&		&\\
		\cline{\sln-18}& \multicolumn{2}{|l|}{Raven~\cite{park2022end}}																&&&\cm&&\cm&&	\multicolumn{1}{c|}{} &&		&&		& \cm &&		&\cm\\
  		\cline{\sln-18}& \multicolumn{2}{|l|}{PostgresML~\cite{postgresml}}																&&\cm&\cm&\cm&\cm&&	\multicolumn{1}{c|}{} &&		&&		& &&		&\\

		\hline
		
		\hline

	\end{tabular}
	}
	\caption{ \smaller Comparison of DB-for-AI/ML systems. (We do not consider approximate co-optimization such as SmartLite~\cite{lin2023smartlite}. SimSQL, SystemML/SystemDS, and TRA are designed for developing AI/ML workloads, lacking direct support for co-optimization of data-oriented queries and AI/ML, and are therefore marked in gray.)
    } 
	\label{relatedWorkMatrix} 
	
\end{figure}

}}

{\color{black}
\subsection{Variation of Running Examples}
\label{sec:variation-examples}

\begin{figure*}[t]
\centering
\includegraphics[width=0.99\textwidth]{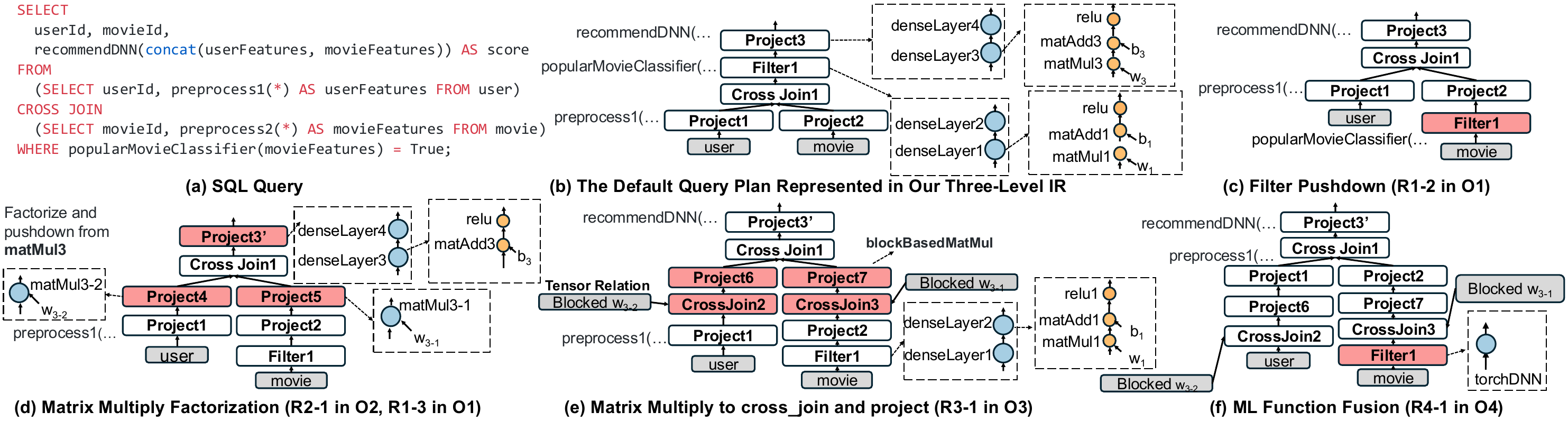}
\caption{\label{fig:examples} \small
\textcolor{black}{A Variant Example of Our Three-Level IR.} \vspace{-15pt}}
\end{figure*}

{In addition to the Running Examples provided in Fig.~\ref{fig:IR-workflow}, Fig.~\ref{fig:running-examples}, and Fig.~\ref{fig:optimizer-workflow}, to illustrate more transformation rules, we provide an variation of the running example in Fig.~\ref{fig:IR-workflow} by changing the two tower recommendation model into DNN-based recommendation model that has two dense layers. In addition, we also added a filter on the movie features, while the filter uses another two-layer DNN model to predict and decide whether the input tuple should be kept or not. Fig.~\ref{fig:examples} uses this variation to illustrate the flexibility and expressiveness of our three-level IR.}
In Fig.~\ref{fig:examples}(a), a query is issued to extract, cross join, and concatenate movie features and user features, and then it applies a DNN model (the \texttt{recommendDNN} function) to each concatenated [user, movie] feature vector to calculate the corresponding recommendation score. The results are further filtered using another DNN model (the \texttt{popularMovieClassifier} function). A default query plan represented in our three-level IR is shown in Fig.~\ref{fig:examples}(b), where the \texttt{Project3} is customized by the \texttt{recommendDNN} function, and \texttt{Filter1} is customized by the \texttt{popularMovieClassifier} function. The details of these functions are described by the middle-level IR consisting of expression operators (i.e., blue nodes). Each opaque expression can be further lowered to bottom-level IR consisting of ML functions (i.e., orange nodes). 
In Fig.~\ref{fig:examples}(c)-(f), we show how our IR supports different co-optimization techniques from O1-O4.

First, pushing down a filter that is customized by a ML function (R1-2 in O1) can be easily performed at the top-level IR, as shown in Fig.~\ref{fig:examples}(c), where \texttt{Filter1} is push down to the \texttt{movie} table without changes to other levels of IRs.

Next, as shown in Fig.~\ref{fig:examples}(d), two optimization techniques are applied. First, it applies R2-1, which factorizes the \texttt{matMul3} from the bottom-level IR of the \texttt{denseLayer3} expression into \texttt{matMul3-1} that multiplies the co-factorized weight matrix $w_{3-1}$ with the preprocessed \texttt{movie} features, and  \texttt{matMul3-2} that performs a similar logic to preprocessed \texttt{user} features. As a result,  the \texttt{Project3} is factorized into \texttt{Project3'}, \texttt{Project4}, and \texttt{Project5}. \texttt{Project3'} corresponds to the rest of the FFNN processing (e.g., the rest of \texttt{denseLayer3} and \texttt{denseLayer4}). \texttt{Project4} and \texttt{Project5}  correspond to the factorized \texttt{matMul3-2} and \texttt{matMul3-1} respectively. They are atomic ML functions and thus they are not lowered to bottom-level IRs. Second, R1-3 is applied to push down both \texttt{Project4} and \texttt{Project5} through the \texttt{CrossJoin1} to avoid redundant computations.

Then, in Fig.~\ref{fig:examples}(e), R3-1 is applied to convert the \texttt{matMul3-1} and \texttt{matMul3-2} into relational operators. In this case, the weight matrices $w_{3-1}$ and $w_{3-2}$ are both vertically partitioned into tiles, e.g., $P3\_1(colId: INT, \boldsymbol{tile}: t \in \mathbb{R}^{d1\times m})$ and $P3\_2(colId: INT, \boldsymbol{tile}: t \in \mathbb{R}^{d2\times m})$, where $d1$ and $d2$ are factorized feature dimension sizes, and $m$ represents the first hidden layer size of the model $recommendDNN$. Therefore, \texttt{matMul3-1} is converted into \texttt{CrossJo-} \texttt{in3} that combines $P3\_1$ and the output relation of \texttt{Project2}, i.e., the preprocessed movie features in the form of $I(movieId: INT, \boldsymbol{V}: x \in \mathbb{R}^{d1})$, where each tuple represents a $d1$-dimensional movie feature vector indexed by $movieId$, followed by \texttt{Project7} that multiplies each joined pair of $t \in 
 P3\_1.tile$ and $x \in R.V$. The transformation of \texttt{matMul3-2} is similar. 

 Finally, as shown in Fig.~\ref{fig:examples}(f), it applies R4-1 to fuse all operators in the \texttt{popularMovieClassifier} model into one high-level ML function that invokes the torchDNN library.
}

\subsection{Weisfeiler-Lehman Encoding Process}
\label{sec:wl-encoding}
{
\color{black}
In this section, we describe how the Weisfeiler-Lehman (WL) kernel is applied to construct positive and negative training samples. The WL kernel encoding procedure is outlined in Algorithm~\ref{alg:wl-feature}. It iteratively generates feature vectors by aggregating and updating node labels. While the encoding processes of Model2Vec and Query2Vec are largely similar, they differ in how the initial node labels are assigned. In Model2Vec, the initial labels are determined by the kernel’s FLOPs values. In contrast, Query2Vec assigns initial labels based on different strategies tailored to each node type. These initialization procedures are detailed in Algorithm~\ref{alg:init-node-label-model2vec} and Algorithm~\ref{alg:init-node-label-query2vec}, respectively.
}

\begin{algorithm}[h]
\footnotesize
\SetKwInOut{Input}{Input}
\SetKwInOut{Output}{Output}
\Input{$root$: root node of the model tree\\
       $T$: number of WL iterations}
\Output{$\mathcal{F}$: WL subtree feature counts}

$\mathcal{L}, \mathcal{H}, \mathcal{N} \gets \emptyset$ \tcp*{Label map, history, node list}

\texttt{DFS}($root, \mathcal{L}, \mathcal{H}, \mathcal{N}$) \tcp*{Initialize the labels through DFS}

\For{$i \gets 1$ \KwTo $T$}{
    $\mathcal{L}_{new} \gets \emptyset$\;
    \ForEach{$node \in \mathcal{N}$}{
        $C \gets \texttt{sorted}([\mathcal{L}[c] \mid c \in node.children])$\;
        $label_{new} \gets \mathcal{L}[node] + C$\;
        $\mathcal{L}_{new}[node] \gets label_{new}$\;
        $\mathcal{H}[node].\texttt{append}(label_{new})$\;
    }
    $\mathcal{L} \gets \mathcal{L}_{new}$\;
}

$\mathcal{F} \gets \emptyset$ \tcp*{Feature vector}
\ForEach{$node \in \mathcal{N}$}{
    \ForEach{$label \in \mathcal{H}[node]$}{
        $\mathcal{F}[label] += 1$\;
    }
}

\Return{$\mathcal{F}$}
\caption{Weisfeiler-Lehman Kernel Feature Extraction}
\label{alg:wl-feature}
\end{algorithm}

\begin{algorithm}[h]
\footnotesize
\SetKwInOut{Input}{Input}
\SetKwInOut{Output}{Output}
\Input{node: a tree node with operator type and FLOPs\\
       $\mathcal{G}$: label groups by kernel type\\
       $\rho$: FLOPs range threshold}
\Output{label: unique label assigned to the node}

$type \gets node.ml\_op\_type$\;
$flops \gets node.ml\_op\_flops$\;
$group \gets \mathcal{G}[type]$\;
$idx \gets \texttt{BinarySearchWithRange}(group, flops, \rho)$\;

\If{$idx = -1$}{
    $label \gets t \_ f$\;
    $group.\texttt{append}(flops)$\;
    $group.\texttt{sort}()$\;
    $\mathcal{G}[type] \gets group$\;
}
\Else{
    $label \gets type +  group[idx]$\;
}
\Return{$label$}
\caption{Initial Node Label Assignment for Model2Vec}
\label{alg:init-node-label-model2vec}
\end{algorithm}

\begin{algorithm}[h]
\footnotesize
\SetKwInOut{Input}{Input}
\SetKwInOut{Output}{Output}
\Input{$root$: root node of the model tree\\
        $\mathcal{L}$: node label map\\
        $\mathcal{H}$: node label history\\
        $\mathcal{N}$: list of all nodes}
\Output{Updated $\mathcal{L}$, $\mathcal{H}$, and $\mathcal{N}$}

\SetKwFunction{FMain}{DFS}
\SetKwProg{Fn}{Function}{:}{}
\Fn{\FMain{$node, \mathcal{L}, \mathcal{H}, \mathcal{N}$}}{
    $label \gets \texttt{InitialNodeLabel}(node)$\;
    $\mathcal{L}[node] \gets label$\;
    $\mathcal{H}[node] \gets [label]$\;
    $\mathcal{N}.\texttt{append}(node)$\;
    \ForEach{$child \in node.children$}{
        \FMain{$child, \mathcal{L}, \mathcal{H}, \mathcal{N}$}\;
    }
}
\caption{DFS Traversal for Label Initialization}
\label{alg:dfs-label-init}
\end{algorithm}

\begin{algorithm}[h]
\footnotesize
\SetKwInOut{Input}{Input}
\SetKwInOut{Output}{Output}
\Input{node: a tree node with type and model embedding \\
       $\mathcal{G}$: label groups by node type \\
       $\mathcal{I}$: FAISS model embedding index \\
       $\tau$: similarity threshold}
\Output{label: label assigned to the node}

$type \gets node.nodeType$\;
$group \gets \mathcal{G}[t]$\;

\If{$node.hashML$}{
    $q \gets normalize(e)$\;
    $idx, sim \gets \mathcal{I}.search(node.mlEmbed)$\;
    \tcp{Assign label based on embedding similarity}
    \If{$sim > \tau$}{
        $label \gets type + idx$
    }
    \Else{
        $label \gets type + next\_idx$\;
        $\mathcal{I}.add(node.mlEmbed)$\;
        $next\_idx \gets next\_idx + 1$\;
    }
}
\tcp{Assign label based on relation data}
\ElseIf{$type = \texttt{TableScanNode}$}{
    $label \gets t + node.table\_id$\;
}
\tcp{Assign label based on filter predicate}
\ElseIf{$type = \texttt{FilterNode}$}{
    $c \gets node.filter["colId"]$\;
    $o \gets node.filter["opId"]$\;
    $v \gets node.filter["val"]$\;
    $label \gets t + c + o + v$\;
}
\tcp{Assign label based on project expression}
\ElseIf{$type = \texttt{ProjectNode}$}{
    $label \gets t$ + node.projectExpr\;
}
\tcp{Assign label based on join predicate}
\ElseIf{$type \in \{\texttt{HashJoinNode}, \texttt{NestedLoopJoinNode}\}$}{
    $label \gets t + node.join$\;
}
\tcp{Assign label based on aggregation keys}
\ElseIf{$type = \texttt{AggregationNode}$}{
    $label \gets type + node.aggKeys$\;
}
\Else{
    $label \gets type$\;
}

$group.add(label)$\;
\Return{$label$}
\caption{Initial Node Label Assignment for Query2Vec}
\label{alg:init-node-label-query2vec}
\end{algorithm}

{\color{black}


\subsection{Error Bound Analysis via $(p,q)$-Consistency}
\label{sec:error-bound-pq}

We model query optimization as a discounted MDP $M=(\mathcal{S},\mathcal{A},P,R,\gamma)$,
where states are query plans and actions are configurable rewrite-rule instantiations.
Let $V^*(s)$ and $Q^*(s,a)$ be optimal values, and define the (root) simple regret
for the input state $s_0$:
\begin{equation}
\label{eq:pq_simpleregret}
\mathcal{R}(s_0)
\;=\;
V^*(s_0) - Q^*(s_0,\hat a)
\;=\;
\max_{a\in\mathcal{A}(s_0)} Q^*(s_0,a) - Q^*(s_0,\hat a),
\end{equation}
where $\hat a$ is the action returned by \CactusDB after $B_{\mathrm{iter}}$ simulations.

The query embedding process $f: \mathcal{S}\rightarrow \mathcal{X}$ is a subjective function mapping ground states $\mathcal{S}$ to their equivalence classes (i.e., abstract states representing clusters of query plans that share the same query embedding) in $\mathcal{X}$. Then, given a logical query plan $s_0$, we view the optimization problem as a Markov Decision Process (MDP) $<\mathcal{T}, \mathcal{A}, \mathcal{P}, \mathcal{R} >$on trajectories, with action space $\mathcal{A}$, state space $\mathcal{T}=\{s_0\}\times (\mathcal{A}\times\mathcal{S})$ being the collection of any state-action sequence of any length that starts in $s_0$ within a length bound $d$. The transition matrix $\mathcal{P}(t, a, t')=\mathcal{P}(s, a, s')$, which represents that the transition probability of $t$ to $t'$ (one step extension of $t$) for $a\in \mathcal{A}$ equals to the transition probability of ground state transitioning $s$ to $s'$ for $a$, where $s$ is the last state in the trajectory $t$, and $s'$ is $t's$ one-step extension in $t'$. We also have $\mathcal{R}(t)$ denoting the reward of selecting $s$ to form trajectory $t$. The Q function and the value function at trajectory $t$ is defined as usual:

\begin{align}
    Q^*(t, a) &= \mathcal{R}(t) + \sum\limits_{t' \in \mathcal{T}}P(t, a, t')V^*(t')\\
    V^*(t) &=\max\limits_{a \in \mathcal{A}}Q^*(t, a)
\end{align}

We then define the application of query embedding $f$ to each ground state $s$ in $t\in \mathcal{T}$, such that $f(t)=f(s_0)a_1f(s_1)\dots a_kf(s_k)$. Then, under $f$, we can define the space of abstracted trajectories at the query embedding level, denoted as $\mathcal{H}=\{h \in \{t_0\} \times (\mathcal{A}\times f): |h| \leq d\}$. Hence, $\mathcal{H}$ forms partition of $\mathcal{T}$.
Then, following existing works on MCTS with state aggregation~\cite{hostetler2014state}, it is easy to prove the following theorem regarding the error bound of the Q function in our proposed optimization process:

\begin{theorem}[{\bf reusable MCTS error bound under $(p,q)$-consistency}]
\label{thm:pq_main_cactus}
Assume our query embedding process $f$ satisfies $(p,q)$-consistency for $p, q \geq 0$ such that for all $h \in \mathcal{H}$,
\begin{align}
\exists &a^* \in \mathcal{A} \;\;:\quad  V^*(t)-Q^*(t,a^*) \le p \quad \forall t \in h
\label{eq:pq_def_p}\\
\quad & |V^*(t_1)-V^*(t_2)| \le q \quad \forall t_1, t_2 \in h.
\label{eq:pq_def_q}
\end{align}
Then, the optimal action in the MCTS tree with state abstraction by $f$, $a^* = \arg\max\limits_{a\in\mathcal{A}}\mathcal{Q}^*(h_0, a)$, has $Q^*$ error in the root state bounded by
\begin{equation}
|\max\limits_{a\in\mathcal{A}} Q^*(s_0, a)-Q^*(s_0, a^*)|\leq 2d(p+q)
\end{equation}
\end{theorem}

\subsection{Proof of Theorem~\ref{thm:pq_main_cactus}}
\label{app:pq_proof}

\subsubsection{Expectimax-tree abstraction and $(p,q)$-consistency}
We briefly restate the abstract expectimax-tree construction (specialized to our notation).
Let $\mathcal{H}$ be the history partition induced by $f$, and let $\mu=\{\mu_h\}$ be a family of distributions over ground trajectories $t\in h$.
The induced abstract-tree transition is
\[
\mathcal{P}_\mu(h,a,h')=\sum_{t\in h}\mu_h(t)\sum_{t'\in h'}\mathcal{P}(t,a,t'),
\]
and the abstract value functions satisfy the standard Bellman optimality equations on the history tree:
\begin{align}
Q^*_\mu(h,a) &= \mathcal{R}_\mu(h) + \sum_{h'\in\mathcal{H}} \mathcal{P}_\mu(h,a,h')\,V^*_\mu(h'),\\
V^*_\mu(h) &= \max_{a\in\mathcal{A}} Q^*_\mu(h,a),
\end{align}
where $\mathcal{R}_\mu(h)=\sum_{t\in h}\mu_h(t)\mathcal{R}(t)$ (see \cite{hostetler2014state}).

We assume $(p,q)$-consistency, i.e., for each $h\in\mathcal{H}$ there exists $a_h^*$ such that
\eqref{eq:pq_def_p}--\eqref{eq:pq_def_q} hold. This matches the definition in \cite{hostetler2014state}.

\subsubsection{Optimal weighting $\mu^*$ and one-step divergence}
In expectimax-tree abstractions, correctness depends not only on $f$ but also on how probability mass is assigned to the ground trajectories inside each history.
Hostetler et al.\ define an \emph{optimal} weighting $\mu^*$ and measure deviation from it via a one-step divergence quantity; we reuse that notion here.
Let $m \triangleq \max_{h\in\mathcal{H}} \delta(\mu,h)\le 1$ be the maximum single-step divergence (notation and formal definition as in \cite{hostetler2014state}).
When reusable MCTS/UCT aggregates statistics by sampling ground transitions and mapping them to histories, it behaves as if it were using $\mu^*$ in the abstract tree (under the conditions discussed in \cite{hostetler2014state}); in the most conservative setting we only assume $m\le 1$.

\subsubsection{Key error decomposition and induction}
Fix any history node $h$ and action $a$. Define the action-value estimation error
\[
E(h,a) \triangleq \Big|Q^*_\mu(h,a) - \sum_{t\in h}\mu_h(t)Q^*(t,a)\Big|.
\]
Following the proof strategy in \cite{hostetler2014state}, decompose the error into (i) error propagated from deeper abstract values and (ii) error introduced by aggregating/weighting at the current step:
\[
E(h,a) \le E_Q(h,a) + E_\delta(h,a),
\]
where $E_Q$ captures the mismatch between $V^*_\mu(\cdot)$ and the weighted ground optimal values below children, and $E_\delta$ captures the single-step weighting/aggregation deviation at the current expansion.

The proof proceeds by induction on the remaining depth $k$ to leaves (base case holds at terminal/absorbing nodes). Assume the inductive hypothesis
\[
E(h',a)\le k(p+m q)\quad\text{for all children }h' \text{ of } h \text{ and all } a\in\mathcal{A}.
\]
Using $(p,\cdot)$-consistency (which allows exchanging $\max$ and $\sum$ with additive penalty $p$) plus the inductive hypothesis yields
\[
E_Q(h,a)\le k(p+m q) + p. 
\]
This corresponds to the step ``exchange max and sum at cost $\le p$'' in \cite{hostetler2014state}.

For the single-step abstraction error, $(\cdot,q)$-consistency implies that within any child history the optimal values vary by at most $q$, and the deviation of $\mu$ from $\mu^*$ scales this by at most the one-step divergence. This yields
\[
E_\delta(h,a)\le m q,
\]
as in the convex-combination argument in \cite{hostetler2014state}.

Combining gives the inductive step:
\[
E(h,a)\le (k+1)(p+m q).
\]
Thus at the root history $h_0$ (depth $d$), for any action $a$ we have $E(h_0,a)\le d(p+m q)$.

\subsubsection{From value-estimation error to root action suboptimality}
Let $a^\sharp\in\arg\max_a Q^*_\mu(h_0,a)$ be the abstract-tree optimal action, and let $a^\star\in\arg\max_a Q^*(s_0,a)$ be the ground-tree optimal root action.
In the worst case, the abstract estimate overvalues $a^\sharp$ by at most $d(p+m q)$ and undervalues $a^\star$ by at most $d(p+m q)$, yielding
\[
Q^*(s_0,a^\star)-Q^*(s_0,a^\sharp) \le 2d(p+m q).
\]
This is exactly the final root-step argument in \cite{hostetler2014state}.
Since $m\le 1$, we also obtain the simplified bound $2d(p+q)$.

\qed

}

{\color{black}
\subsection{GPU Acceleration}

\noindent
\textbf{Further discussion: GPU Support} 
While this work focuses on CPU, our reusable MCTS optimizer is aware of the equivalence between CPU and GPU implementation of the same ML function and leverages R4-2 to invoke LibTorch CUDA library~\cite{libtorch_cuda} in our GPU environment (see the beginning of Sec.~\ref{sec:evaluation}). As shown in Tab.~\ref{tab:ml-gpu}, GPU offloading achieved by our optimizer yields an additional $17\%-55\%$ performance improvement over the optimized query plan and accelerates model inference by $1.3\times$ on average.
However, fully optimizing GPU acceleration raises challenges in pipelining, load balancing, and CPU–GPU coordination. These aspects are beyond the scope of this paper and are left for future work.

\begin{table}[H]
\color{black}
\tiny
\centering
\caption{\small Recommendation Query Performance with GPU Enabled}
\label{tab:ml-gpu}
\resizebox{\columnwidth}{!}{
\begin{tabular}{|c|c|c|c|cc|}
\hline
\multirow{2}{*}{} & \multirow{2}{*}{EvaDB} & \multirow{2}{*}{Apache PySpark w/UDF} & \multirow{2}{*}{DL-Centric} & \multicolumn{2}{c|}{CactusDB}                             \\ \cline{5-6} 
                  &                        &                                       &                             & \multicolumn{1}{c|}{w/o GPU Extension} & w/ GPU Extension \\ \hline
Q1                &        40.40 sec                &            48.56 sec                           &           47.2 sec                  & \multicolumn{1}{c|}{0.14 sec}          & \textbf{0.09 sec}         \\ \hline
Q2                &          Failed (OOM)              &                Failed (OOM)                       &            119.54 sec                 & \multicolumn{1}{c|}{2.71 sec}                  &    \textbf{2.14 sec}              \\ \hline
Q3                &         Failed (OOM)               &               Failed (OOM)                      &                172.39 sec             &  \multicolumn{1}{c|}{18.67 sec}         & \textbf{15.93 sec}        \\ \hline
\end{tabular}
}
\end{table}
}

{\color{black}
\subsection{Extended Related Works} 
\label{sec:extended-related-works}
\noindent
\textbf{In-DB AI/ML Systems.} To the best of our knowledge, no existing system fully supports all categories of SQL-ML co-optimization techniques (O1–O4).
UDF-centric systems such as EvaDB~\cite{kakkar2023eva} and PostgresML~\cite{postgresml} support most O1 techniques by allowing inference logic to be encapsulated in user-defined functions. MADLib~\cite{hellerstein2012madlib} and PySpark~\cite{singh2022manage} provide partial support for O1, but handle arbitrary UDFs poorly. For instance, MADLib cannot push down table-level UDFs like \textit{madlib.mlp\_classification}, and PySpark only pushes down MLlib functions and not arbitrary UDFs. These systems cannot analyze or transform UDF internals, making them unsuitable for deeper co-optimization as required by O2–O4.
Existing factorized ML systems~\cite{factorize-norm-data,factorize-lmfao, factorize-joinboost, factorize-la, li2019enabling, schleich2016learning} focus on ML training processes, relying on predefined factorizable AI/ML models or linear algebra (LA) operators. Although related to O2-style optimization, they lack dynamic decomposition of arbitrary inference pipelines and integration with O3/O4 type optimizations.
\textcolor{black}{IMBridge~\cite{zhang2024imbridge, zhang2025mitigating} optimizes query performance through automated ML inference context caching and batch inference. Nonetheless, it lacks support for O2 and O3, which are essential for comprehensive end-to-end performance optimization.}
Other systems like SimSQL~\cite{luo2018scalable,jankov12declarative}, SystemML~\cite{boehm2016systemml}, and SystemDS~\cite{boehm2019systemds} embed ML workloads in relational algebra atop big data processing platforms like Hadoop~\cite{white2012hadoop} or Spark~\cite{zaharia2010spark}. MASQ~\cite{paganelli2023pushing,del2021transforming} rewrites limited inference operators (e.g., one-hot encoding, linear models, decision trees) into SQL. None explore partial translation of LA operators into relational form for O2/O4 co-optimization.

\noindent
\textbf{Intermediate Representations (IRs).}
Several unified IRs have been proposed for SQL-ML co-optimization, though most cover only subsets of O1–O4.
Raven~\cite{park2022end} combines relational, LA, and ML operators, supporting O1 and O4 and partially O3 by converting full ML pipelines into relational form, reducing flexibility. 
SDQL~\cite{shaikhha2022functional} uses a semi-ring IR covering O1, O2, and O4, but cannot transform LA into relational operators (lacking O3).
Weld~\cite{palkar2017weld} 
unifies computation across libraries through vector operators (e.g., \texttt{vecbuilder}, \texttt{vecmerger}), enabling fusion and data-movement optimization, yet most O1–O3 techniques must be implemented outside of the IR~\cite{palkar2018evaluating}.
%
Lara~\cite{kunft2019intermediate} uses monad comprehensions for UDFs, enabling O1/O4 optimizations (push-down, layout tuning) but not O2/O3.
LingoDB~\cite{jungmair2022designing,jungmair2023declarative} leverages MLIR for combining relational and non-relational workloads (e.g., $k$-means, PageRank)~\cite{lattner2020mlir}. While flexible, it does not target SQL-ML co-optimization techniques in O2–O4. Tensor Relational Algebra~\cite{DBLP:journals/pvldb/YuanJZTBJ21} primarily addresses O3-style integration of tensors with relational queries, without broader support for O1, O2, or O4.

\vspace{3pt}
\noindent
\textbf{Inference Query Optimization.}
Raven~\cite{park2022end} applies logical optimizations in a fixed order using rule-based, classification, and regression strategies to choose among ML2SQL, ML2DNN, or no optimization.  Most other in-database AI/ML systems rely on rules and heuristics for inference query optimization. 
Weld~\cite{palkar2017weld} and Lara~\cite{kunft2019intermediate} use rule-based plan selection, while SDQL [48] applies rewrite rules until reaching a fixed point.


 Database query optimization has evolved into a multifaceted research domain, encompassing rule-based, cost-based, and learning-based approaches. 
 \textbf{Rule-based optimizers (RBOs)} apply predefined transformations (e.g., Spark Catalyst ~\cite{pirahesh1992extensible, armbrust2015spark}), but fixed ordering can yield suboptimal plans.
 \textbf{Cost-Based Optimizer (CBO)}~\cite{graefe1995cascades, calcite, postgresqldb} estimates the cost of query execution using statistical models and employs techniques such as dynamic programming, genetic algorithms, and Monte Carlo Tree Search (MCTS)~\cite{zhou2021learned, sikdar2020monsoon}. 
 \textbf{Learned optimizers (LBOs)} ~\cite{gaussml, zhou2021learned, marcus2019neo, marcus2021bao} use ML for cost or cardinality estimation but struggle with ad-hoc AI/ML queries unseen during training.

\vspace{5pt}
\noindent
\textbf{Our Work.}
To close the gaps, we proposed \CactusDB, a novel in-database AI/ML system focusing on inference queries. 
Like Raven~\cite{park2022end}, and different from Weld~\cite{palkar2017weld}, Lara~\cite{kunft2019intermediate}, LingoDB~\cite{jungmair2023declarative}, and SDQL~\cite{shaikhha2022functional}, \CactusDB allows relational, linear-algebra, and ML operators to co-exist for expressive and flexible query processing. But our system is distinct from Raven in the following aspects:
(1) Unlike Raven’s ONNX-based IR, \CactusDB encodes AI/ML semantics using native database expressions in UDF-centric systems, forming a three-level IR. This yields tighter integration with existing optimizers, improved partial evaluation, and seamless adoption in SQL/UDF systems.
(2) Instead of treating ML pipelines as monolithic graphs, \CactusDB decomposes inference pipelines into fine-grained sub-computations, enabling selective optimizations such as factorization, push-down, and tensor-relational conversion.
(3) While Raven explores a small rule-based search space (ML2SQL, ML2DNN, or none), \CactusDB introduces a new optimizer combining query embeddings with MCTS to efficiently search the exponentially large space of co-optimization strategies.
}

\subsection{Vanilla MCTS Algorithm}
The Alg.~\ref{alg:mcts} describes how the vanilla MCTS algorithm is used in query optimization.
\begin{algorithm}[htbp]
\footnotesize
\SetKwInOut{Input}{Input}
\Input{ $root$, the state of the original query plan; 
$B_{iteration}$, the budget of MCTS training iterations}
\For {i in range($B_{iteration}$)}{
    $node$ $\gets$ $root$ \;
    \tcc{If the current node is fully expanded, perform node selection. Otherwise, perform node expansion.}
    \While{node.isTerminalState() == False}{
        \If{node.expanded}{
            $node$ $\gets$ $select(node)$ \;
        }
        \Else{
            $node$ $\gets$ $expand(node)$ \;
            $node'$ $\gets$ $rollout(node)$ \;
            $\textbf{break}$;
        }
    }
    $reward$ $\gets$ $computeReward(node')$ \;
    $backpropagate(node, reward)$;
}
$\textbf{return}$;
\caption{General MCTS Algorithm}
\label{alg:mcts}
\end{algorithm}

\subsection{Recommendation Queries}
\label{sec:movielens}
\begin{figure}[ht]
\centering
\includegraphics[width=0.4\textwidth]{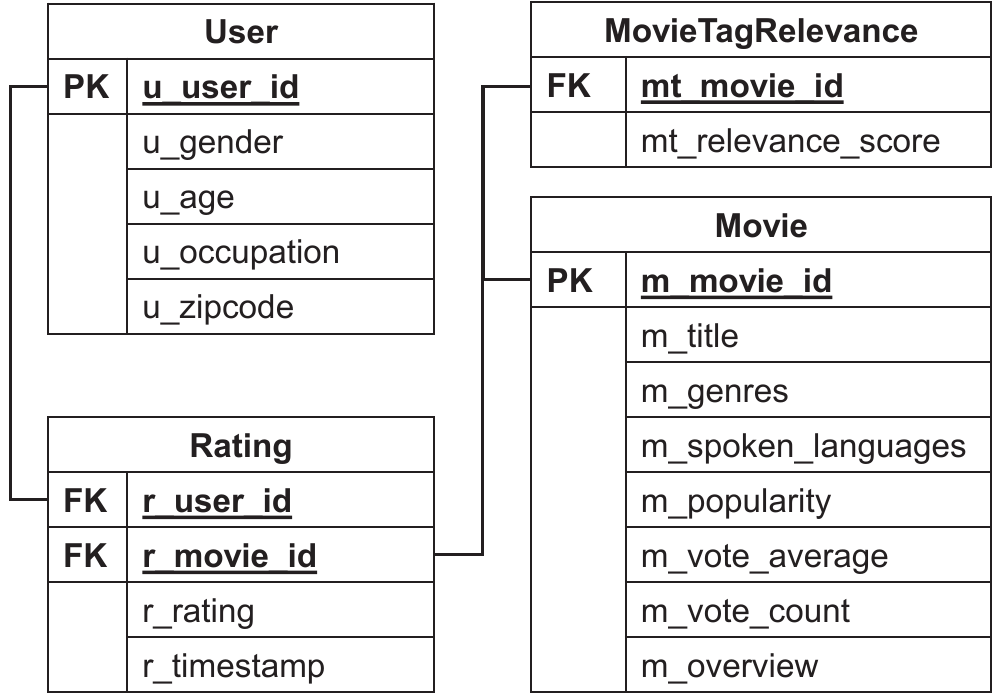}
\caption{ER Diagram of Movielens Database}
\label{fig:ml-er}
\end{figure}
The three recommendation queries we used in evaluation are described in List.~\ref{lst:movielens-q1} based on the MovieLens dataset, with the corresponding ER diagram shown in Fig.~\ref{fig:ml-er}. 

In Q1, the trending model has an input size of $3$, with hidden layers of $128$ and $64$ units. ReLU is used as the activation function. The output layer consists of a single neuron with a sigmoid function. The two tower model initialized embedding tables of dimension $32$ for each categorical feature. Later, the concatenated user and movie features are then passed to a fully connected neural network with the two hidden layers, each containing $300$ neurons. The output layer consists of $128$ neurons. The input features dimensions are $129$ for each user and $65$ for each movie. 

In Q2, the trending model is the same as the one used in Q1. The \texttt{predict\_user\_interest\_DNN} is a fully connected neural network with an input dimension of $259$ and a single hidden layer containing $128$ neurons. Its output consists of two neurons with a softmax activation function. The Autoencoder model takes an input of size $140,979$, representing the number of unique movie tags, with a hidden layer of $2,048$ neurons. The output is a dense layer with $256$ neurons. The DLRM model initializes embedding tables of dimension $128$ for each categorical variable. The numerical features are processed by a one-layer MLP, a fully connected neural network with an input size of $256$ and output size of $128$. The categorical features' embedding vectors are concatenated with the MLP's output and then passed to another one-layer MLP, with an input size of $256$, a hidden layer of $128$ neurons, and a single-neuron output layer.

In Q3, \texttt{autoencoder} and \texttt{predict\_user\_interest\_DNN} are both fully connected neural networks with the same architecture as described in Q2. \texttt{predict\_rating\_DNN} is another fully connected neural network, which has an input dimension of $5$, two hidden layers with $512$ and $1024$ neurons, respectively, and an output layer of size $6$.

\begin{lstlisting}[style=sqlstyle,caption={Query 1}, label={lst:movielens-q1}] 
// Q1
SELECT
 two_tower_model (user_feature, movie_feature)
FROM
 (
   SELECT
     concat(embedding (user_id), embedding (gender), embedding (age), embedding (occupation), user_avg_rating) AS user_feature
   FROM
     USER u
     JOIN (
       SELECT user_id, average (rating) AS user_avg_rating
       FROM rating GROUP BY user_id
     ) ON u.user_id=user_id
 )
 CROSS JOIN (
   SELECT
     concat(embedding (movie_id), embedding (genres), movie_avg_rating) AS movie_feature
   FROM
     movie m
     JOIN (
       SELECT movie_id, average (rating)
       FROM rating GROUP BY movie_id
     ) ON m.movie_id=movie_id
   WHERE
     movie.genres LIKE '%Action%' AND trending_movie_DNN(movie_feature)=1
 )

// Q2
SELECT
  DLRM(user_feature,movie_feature,movie_dense_feature)
FROM
  (SELECT
      feature_extraction1(*) AS user_feature
    FROM
      user
      CROSS JOIN (
        SELECT
          feature_extraction2(movie_feature),
          autoencoder(movie_relevance_tag) AS movie_dense_feature
        FROM
          movie
        JOIN movie_tag_feature ON movie.m_movie_id=movie_tag_feature.mt_movie_id
      )
    WHERE
      predict_trending_movie_DNN(movie_feature)=1
      AND predict_user_interest_DNN(user_feature, movie_dense_feature)=1)


// Q3
SELECT
  cosine_similarity(movie_dense_feature1, movie_dense_feature2) AS relevant_score
FROM
  (
    SELECT
      autoencoder(mt_relevance_score) AS movie_dense_feature1
    FROM
      (
        SELECT
          preprocess1(user.*) AS user_feature,
          preprocess2(movie.*) AS movie_feature
        FROM
          user
          CROSS JOIN movie
          JOIN movie_tag_relevance ON mt.movie_id = m_movie_id
      )
    WHERE
      genres LIKE '%Fiction%'
      AND predict_user_interest_DNN(user_feature, movie_feature) = 1
      AND predict_rating_DNN(user_feature, movie_feature) > 3
  )
  CROSS JOIN(
    SELECT
      autoencoder(mt_relevance_score) AS movie_dense_feature2
    FROM
      movie_tag_relevance
  )


\end{lstlisting}

\subsection{Retailing-Complex Queries}
\label{appex:tpcxai-qeuries}
\begin{figure}[ht]
\centering
\includegraphics[width=0.5\textwidth]{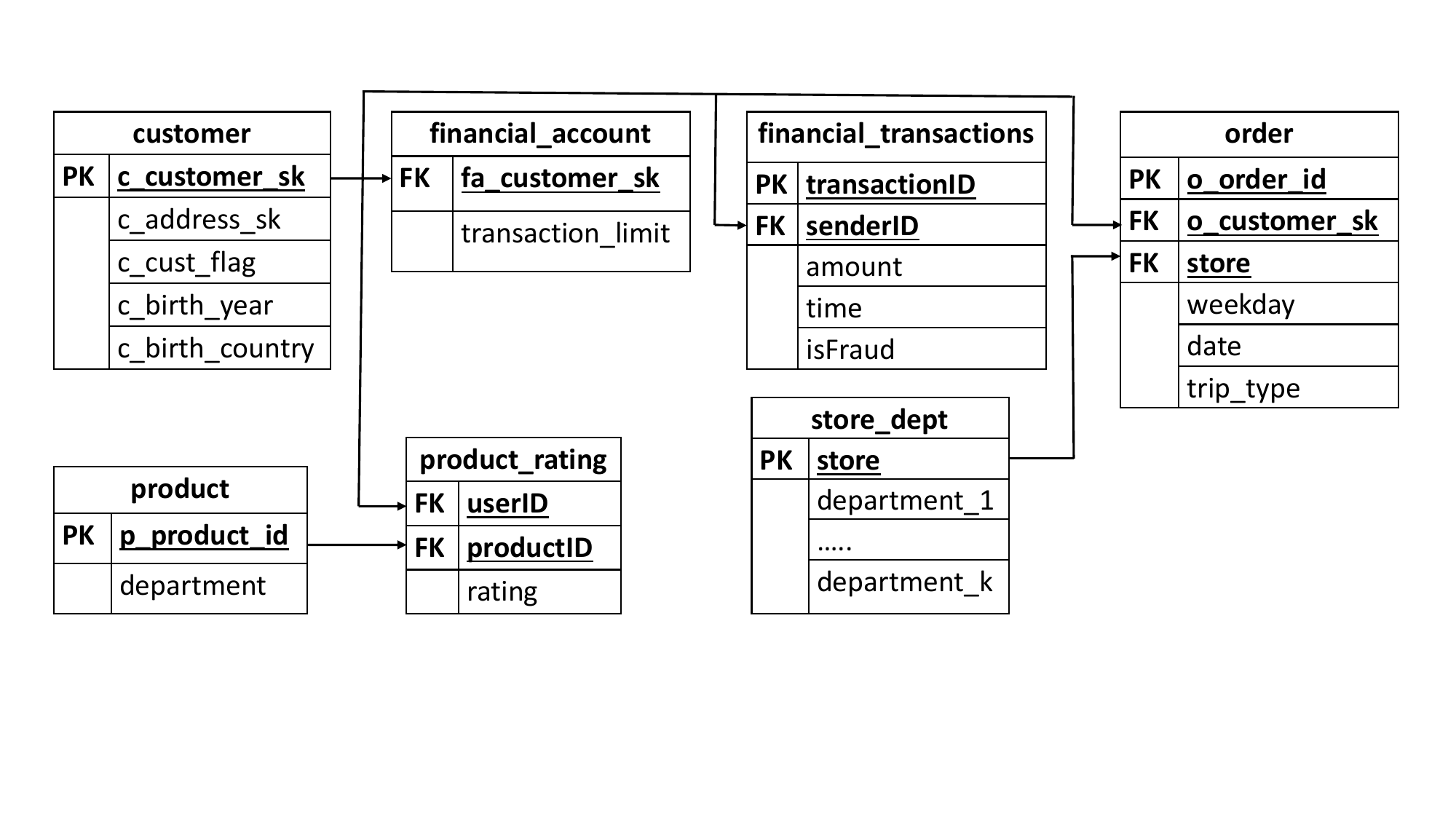}
\caption{Simplified ER Diagram of Retailing Dataset from TPCx-AI~\cite{tpcx-ai}}
\label{fig:tpcx-ai-er}
\end{figure}

\noindent
List.~\ref{lst:tpcxai-q} presents three complex queries based on the TPCx-AI dataset, with the corresponding ER diagram shown in Fig.~\ref{fig:tpcx-ai-er}. 

In Q1, \texttt{is\_popular\_store} is a UDF that computes a popularity score for each store based on its department availability. \texttt{trip\_class} \texttt{ifier\_dnn} is a neural network comprising a customer embedding table and a fully connected neural network. The embedding table is initialized with the dimension of $16$, and the resulting customer embedding vector is concatenated with additional features, forming an input of size $63$. The input is then passed through a fully connected neural network with two hidden layers of $48$ and $24$ neurons, respectively, followed by an output layer of size $1000$, representing the trip types. 

In Q2, \texttt{dnn\_fraud\_predict} is a deep neural network (DNN) model designed for fraud classification, consisting of an input layer of size $32$, a single hidden layer with $12$ neurons, and an output layer of size $2$. \texttt{xgboost\_fraud\_predict} is an XGBoost model with $1,600$ trees which take $5$ features. 

Q3 invokes a two-tower model that initializes multiple embedding tables for categorical variables, each with a dimension of either $16$ or $8$. Customer features and product features are processed separately through the user tower and product tower. The user tower is a fully connected neural network with an input size of $50$, followed by two hidden layers with $128$ and $64$ neurons and an output layer of $16$ neurons. The product tower has a similar structure but with a different input size of $25$.

\begin{lstlisting}[style=sqlstyle,caption={Queries on Retailing Dataset}, label={lst:tpcxai-q}] 
// Q1
SELECT o_order_id, trip_classifier_dnn (concat (order_feature, store_feature))
FROM (
  SELECT o_order_id , get_order_features(o_customer_sk, weekday, date) AS order_feature, store_feature
  FROM order o
  JOIN (
    SELECT store, concat (*) AS store_feature
    FROM store_dept_wide s
  ) ON o. store  = s.store
  WHERE o. weekday != 'Sunday' AND is_popular_store (store_feature) = 1
)


// Q2
SELECT transactionID
FROM (
  SELECT c_customer_sk , get_customer_feature (c_address_sk, c_cust_flag, c_birth_year, c_birth_country, transaction_limit) AS customer_feature
  FROM customer
  JOIN (
    SELECT fa_customer_sk, transaction_limit
    FROM financial_account
  ) ON c_customer_sk  = fa_customer_sk
  WHERE c_cust_flag = 0
  JOIN (
    SELECT transactionID, senderID, get_transaction_feature (*) AS transaction_feature
    FROM financial_transactions t
  ) ON c_customer_sk  = senderID
  WHERE age_during_transaction ( c_birth_year , t.time) >= 18 AND is_working_day (t.time) = 1
)
WHERE xgboost_fraud_predict(concat (customer_feature, transaction_feature)) >= 0.5 AND dnn_fraud_predict(concat (customer_feature, transaction_feature)) = 1



// Q3
SELECT two_tower_model (customer_feature, product_feature)
FROM (
  SELECT concat(embedding (c_customer_sk), embedding (c_address_sk), c_cust_flag, embedding (get_age (c_birth_year)),    embedding (birth_country), customer_avg_rating) AS customer_feature
  FROM customer c
  JOIN (
    SELECT userID, AVG (rating) AS customer_avg_rating
    FROM product_rating GROUP BY userID
  ) AS cr ON c. c_customer_sk = cr.userID
) AS customer_outer
CROSS JOIN (
  SELECT concat(embedding (p_product_id), embedding(department), product_avg_rating) AS product_feature
  FROM product p
  JOIN (
    SELECT productID, AVG (rating) AS prod_avg_rating
    FROM product_rating GROUP BY productID
  ) AS pr ON p. p_product_id = pr.productID
  WHERE pr.prod_avg_rating >= 4.0
) AS prod_outer

\end{lstlisting}

\subsection{LLM Queries}
\label{sec:appendix-llm}
List.~\ref{lst:llm-query} describes the two LLM queries used in our evaluation. We utilize OpenAI's RESTful API to invoke the \texttt{gpt-3.5-turbo} model to summarize the user/movie information and make the recommendations. The \texttt{trending\_movie\_classifier} is a fully connected neural network with the same structure as the one described in Sec.~\ref{sec:movielens}. Q2 employs a RAG-based approach to retrieve movies' relevant information by performing a nearest-neighbor vector search on pre-built document embedding vectors using Faiss's cosine similarity index.
\begin{lstlisting}[style=sqlstyle,caption={LLM Queries}, label={lst:llm-query}] 
// Q1: Using LLM for Movie Recommendation 
SELECT LLM('Please give a recommendation score and explain why', LLM('Please summarize', u.description), LLM('Please summarize', m.description))
FROM user u CROSS JOIN movie m
WHERE m.spoken_language LIKE % English % AND trending_movie_classifier(m.popularity, m.vote_average, m.vote_num) = True;

// Q2: Using RAG to retrieve relevant information with given movie title and using LLM for recommendation
SELECT LLM('Please give a recommendation score and explain why', LLM('Please summarize', u.description), RAG(m.title))
FROM user u CROSS JOIN movie m
WHERE m.spoken_language LIKE % English % AND trending_movie_classifier(m.popularity, m.vote_average, m.vote_num) = True;
\end{lstlisting}

{
\color{black}
\begin{figure}[ht]
\centering
\includegraphics[width=0.45\textwidth]{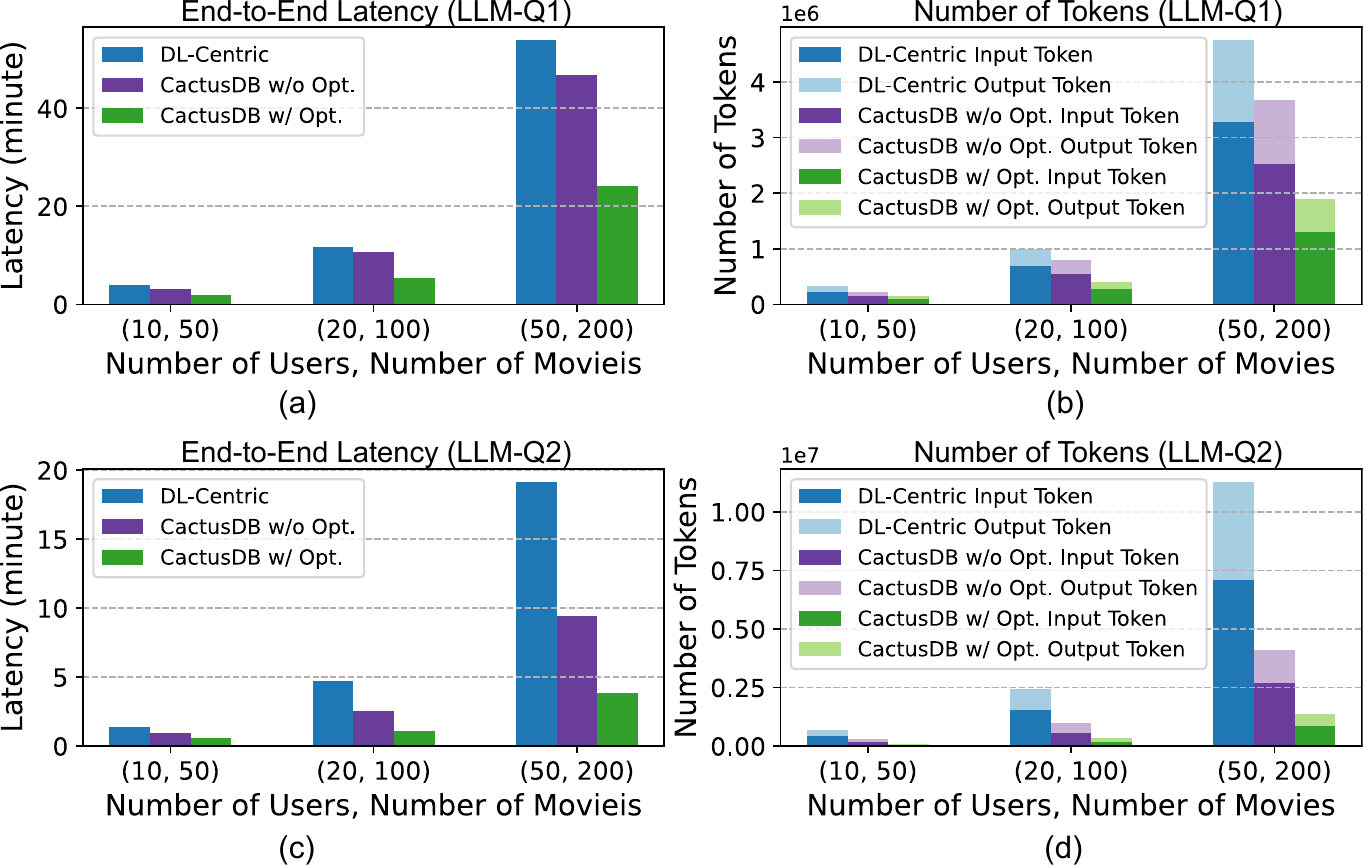}
\caption{\small Evaluating the LLM Queries with R4-1 and R1-3 
}
\label{fig:eval-llm}
\end{figure}


\noindent
\textbf{LLM Inference Queries.} In this benchmark, the overheads of LLM inferences dominate the end-to-end latency. In addition, the monetary costs of LLM usage linearly increase with the number of tokens sent to and received from LLM. Q2 replaces the LLM-based movie summarization with RAG, which results in faster execution but increases the number of tokens compared to Q1. As shown in Fig.~\ref{fig:eval-llm}, \textbf{our \CactusDB is capable of factorizing the nested LLM functions and pushing down the LLM summarization call and RAG call before the \texttt{cross\_join} to avoid repeatedly LLM inferences on the same records, resulting in $3\times$ speed up and $72.4\%$ reduction of tokens (and costs) on average.} We noticed that even without enabling our reusable MCTS query optimizer, \CactusDB benefits from Velox's built-in runtime optimization, such as filter push-down. These runtime optimizations effectively reduced the total number of input/output tokens sent to the LLMs by $25\%$ for Q1 and $59\%$ for Q2 on average with end-to-end latency speedups of $1.1\times$ to $2.1\times$.  

}
\subsection{Comparison of Optimization Algorithms on Recommendation Queries}

\begin{table}[h]
\begin{center}
\scriptsize
\caption{\small Comparison of Optimization Algorithms (unit: second)}
\label{tab:optimier-exp-ml}
\resizebox{\columnwidth}{!}{
\begin{tabular}{|>{\centering\arraybackslash}p{1.47cm}|>{\centering\arraybackslash}p{0.35cm}|>{\centering\arraybackslash}p{0.35cm}|>{\centering\arraybackslash}p{0.35cm}|>{\centering\arraybackslash}p{0.35cm}|>{\centering\arraybackslash}p{0.35cm}|>{\centering\arraybackslash}p{0.35cm}|>{\centering\arraybackslash}p{0.35cm}|>{\centering\arraybackslash}p{0.35cm}|>{\centering\arraybackslash}p{0.35cm}|}
\hline
 &\multicolumn{3}{c|}{Recommendation-Q1} & \multicolumn{3}{c|}{Recommendation-Q2} & \multicolumn{3}{c|}{Recommendation-Q3}\\
&Opt.&Exec.&Total&Opt.&Exec.&Total&Opt.&Exec.&Total\\ 
\hline
\hline
Un-optimized&  NA                         & 144.2                                       & 144.2&  NA                         & 13.6                                        & 13.6&  NA                         & 36.5                                       & 36.5\\
Arbitrary& 0.9                                            & 1.8                                         & 2.7& 5.1                                            & 131.3                                       & 136.3& 4.4                                            & 23.7                                        & 28.1\\
Heuristic& \textbf{0.2}                                            & 5.2                                         & 5.4& \textbf{0.3}                                            & 12.3                                        & 12.6& \textbf{0.1}                                            & 16.3                                        & 16.3\\
Vanilla MCTS& 5.2                                            & 0.3                                         & 5.5& 6.8                                            & 6.8                                         & 13.6& 10.3                                           & \textbf{8.7}                                         & 18.9\\
\textbf{Reusable MCTS}& 1.5                                            & \textbf{0.2}                                         & \textbf{1.7}& 2.1                                            & \textbf{6.5}                                         & \textbf{8.6}& 3.6                                          & 8.8                                         & \textbf{12.4}\\
\hline
\end{tabular}
}
\end{center}
\end{table}

\noindent
\textbf{Recommendation queries.} We break down the end-to-end latency of each query into two parts: query optimization latency and query execution latency. 
Tab.~\ref{tab:optimier-exp-ml} compares the latency for query optimization (Opt.), query execution (Exec.), and the total of them. \textbf{Our reusable MCTS strikes the best trade-off between overhead and effectiveness compared to other baselines. Even though it introduces additional overhead for query embedding, the query optimization latency is greatly saved by reusing the MCTS.} The arbitrary optimization strategy could lead to better or worse query execution time compared to unoptimized queries. That is because not all optimization rules will be beneficial. Heuristic optimization strategy will always lead to better efficiency (optimization latency). However, simply relying on pre-defined heuristics cannot exploit the best optimization techniques for all queries. Vanilla MCTS demonstrates effectiveness in searching for optimal query plans due to its query-specific optimization. However, it requires building an MCTS search tree from scratch every time, which worsens the optimization latency. 

\subsection{Synthetic Model Sampling}
\label{sec:app-model-sampling}
{
\color{black}

We design a model sampling strategy to generate diverse architectures for collecting the training data for the Model2Vec model. The strategy is built on a set of reusable model-building blocks, which we denote as parameterized operators:

\begin{itemize}
  \item \texttt{ConvolutionalLayer} (in, out, kernel, stride, padding)
  \item \texttt{PoolingLayer} (kernel, stride, padding)
  \item \texttt{LinearLayer} (in, out, activation, bias)
  \item \texttt{BatchNormalization} (num\_features, $\epsilon$)
  \item \texttt{Dropout}(p)
  \item \texttt{EmbeddingLayer} (num\_tuples, dim)
  \item \texttt{SVD} (dim)
  \item \texttt{Decision\_Forest} ($n_{trees}$, $d_{max}$)
  \item \texttt{DecisionTree} ($d_{max}$)
  \item \texttt{Activation} $\in \{\text{ReLU}, \text{Sigmoid}, \text{Tanh}, \text{Softmax}\}$
  \item \texttt{Concat} ($\cdot$)
  \item \texttt{CosineSimilarity} ($\cdot, \cdot$)
  \item \texttt{Flatten} ($\cdot$)
\end{itemize}

Each operator can be parameterized within a configurable range, and blocks may be repeated between a minimum and maximum depth.
Using these primitives, users can compose customized model templates based on their needs. In our work, we define a set of representative model templates, which are used to generate training data, as illustrated in List.~\ref{lst:model-templates}:

\begin{lstlisting}[style=sqlstyle,caption={Predefined model templates},label={lst:model-templates}]
// Multi-Layer Perceptron (MLP)
MLP :=
  Repeat{min_repeat, max_repeat}(
    LinearLayer(RANGE(in_min, in_max),
                RANGE(out_min, out_max),
                CHOICE(Activation))
  )

// Two-Tower Model
Tower :=
  Repeat{min_repeat, max_repeat}(
    LinearLayer(RANGE(in_min, in_max),
                RANGE(out_min, out_max),
                CHOICE(Activation))
  )
TwoTowerModel := CosineSimilarity(Tower, Tower)

// Deep Learning Recommendation Model (DLRM-style)
DLRM :=
  Concat(
    Repeat{min_repeat, max_repeat}(
      EmbeddingLayer(RANGE(num_min, num_max),
                     RANGE(dim_min, dim_max))
    ),
    Repeat{min_repeat, max_repeat}(
      LinearLayer(RANGE(in_min, in_max),
                  RANGE(out_min, out_max),
                  CHOICE(Activation))
    )
  )
  -> Repeat{min_repeat, max_repeat}(
       LinearLayer(RANGE(in_min, in_max),
                   RANGE(out_min, out_max),
                   CHOICE(Activation))
     )

// Convolutional Neural Network (CNN)
CNN :=
  Repeat{min_repeat, max_repeat}(
    ConvolutionalLayer(RANGE(in_min, in_max),
                       RANGE(out_min, out_max),
                       RANGE(kernel_min, kernel_max))
    -> BatchNorm(num_features)
    -> CHOICE(Activation)
    -> PoolingLayer(RANGE(kernel_min, kernel_max))
  )
  -> Flatten()
  -> Repeat{min_repeat, max_repeat}(
       LinearLayer(RANGE(in_min, in_max),
                   RANGE(out_min, out_max),
                   CHOICE(Activation))
     )

// Decision Forest
DecisionForest :=
  DecisionForest(RANGE(n_trees_min, n_trees_max),
                 RANGE(depth_min, depth_max))
\end{lstlisting}

In practice, when sampling a model, the system randomly selects one of these templates and instantiates its building blocks by drawing parameter values (e.g., layer widths, kernel sizes, embedding dimensions, number of trees) from the specified ranges. This process yields a rich variety of architectures, while ensuring that the sampled models remain consistent with patterns observed in real-world in-database ML workloads.
}

\subsection{Synthetic Query Templates}
\label{sec:app-query-sampling}

{
\color{black}
We describe the query templates used to generate training data and evaluation queries. In the MovieLens dataset, we convert the three queries described in Sec.~\ref{sec:movielens} into templates and pick the seven most voted data science pipelines from Kaggle~\cite{movielens2020kaggle}. In the TPCx-AI~\cite{brucke2023tpcx} dataset, in addition to the tree queries we described in Sec~\ref{appex:tpcxai-qeuries}, we convert the other use cases queries into templates as illustrated in List.~\ref{lst:movielens-templates} and List.~\ref{lst:tpcxai-templates}. The query sampler randomly samples a model architecture and configuration parameters, and randomly samples filters with different selectivity. 
}


\begin{lstlisting}[style=sqlstyle,caption={Query templates for MovieLens dataset}, label={lst:movielens-templates}] 
// Template 4: User Rating Prediction for Movies
SELECT DNN(process_gender(gender), normalize_age(age), normalize_occupation(occupation), split_genres(genres)) 
FROM user CROSS JOIN movie
[WHERE AGE_FILTER | GENDER_FILTER | OCCUPATION_FILTER | GENRE_FILTER];

// Template 5: User Opinion Prediction (Negative/Neutral/Positive)
SELECT DNN(user_id, movie_id) 
FROM user 
[WHERE AGE_FILTER | GENDER_FILTER | OCCUPATION_FILTER | ZIP_CODE_FILTER];

// Template 6: Recommend Movies to Users via SVD
SELECT SVD(user_id, movie_id) 
FROM user CROSS JOIN movie 
[WHERE AGE_FILTER | GENDER_FILTER | OCCUPATION_FILTER | ZIP_CODE_FILTER | GENRE_FILTER];

// Template 7: Rating Prediction with Collaborative Filtering
SELECT LightFM(user_id, movie_id) 
FROM user CROSS JOIN movie 
[WHERE AGE_FILTER | GENDER_FILTER | OCCUPATION_FILTER | ZIP_CODE_FILTER | GENRE_FILTER];

// Template 8: Rating Prediction with AutoEncoder
SELECT AutoEncoder(user_id, movie_id) 
FROM user 
[WHERE AGE_FILTER | GENDER_FILTER | OCCUPATION_FILTER | ZIP_CODE_FILTER];

// Template 9: Detect Gender Sterotypes with DNN
SELECT DNN(prepare_feature(...)) 
FROM ratings JOIN movie ON ratings.movie_id = movie.movie_id 
[WHERE GENRE_FILTER | TIMESTAMP_FILTER];

// Template 10: Rating Prediction
SELECT DNN(movie_id, age, occupation) 
FROM user CROSS JOIN movie 
[WHERE AGE_FILTER | GENDER_FILTER | OCCUPATION_FILTER | ZIP_CODE_FILTER | GENRE_FILTER];
\end{lstlisting}

\begin{lstlisting}[style=sqlstyle,caption={Query templates For TPCx-AI dataset}, label={lst:tpcxai-templates}]
// Template 4: Product Rating Prediction with SVD
SELECT SVD(user_id, product_id) 
FROM product_rating 
JOIN product ON productID = p_product_id 
JOIN customer ON c_customer_sk = userID 
[WHERE PRODUCT_DEPARTMENT_FILTER | CUSTOMER_BIRTH_DATE_FILTER | CUSTOMER_BIRTH_COUNTRY_FILTER];

// Template 5: Spam Review Detection with DNN
SELECT DNN(tf_idf_features(text)) 
FROM review_comments 
[WHERE ID_FILTER];

// Template 6: Classification of Trips with DNN/Decision Forest
SELECT DNN/DecisionForest(prepare_feature(...)) 
FROM (
  SELECT o_order_id, department, quantity,
         SUM(quantity) AS scan_count,
         MIN(EXTRACT(DOW FROM date)) AS weekday
  FROM tpcxai_order_serving
  JOIN tpcxai_lineitem_serving ON o_order_id = li_order_id
  JOIN tpcxai_product_serving ON li_product_id = p_product_id
  GROUP BY o_order_id, date, department, quantity
)
[WHERE TIMESTAMP_FILTER | DEPARTMENT_FILTER | WEEKDAY_FILTER | PRICE_FILTER | QUANTITY_FILTER];

// Template 7: Fraud Detection
SELECT DNN/LogisticRegression(concat(EXTRACT(HOUR FROM time)/23, amount/transaction_limit)) 
FROM financial_transactions 
JOIN financial_account ON sender_id = fa_customer_sk 
JOIN customer ON fa_customer_sk = c_customer_sk 
[WHERE TIMESTAMP_FILTER | AMOUNT_FILTER | CUSTOMER_BIRTH_INFO_FILTER];

// Template 8: Sales Prediction
SELECT DNN(store_id_encoder(store_id), department_encoder(department), normalize_week(week)) 
FROM store_dept 
[WHERE DEPARTMENT_FILTER | WEEK_FILTER | STORE_FILTER];

// Template 9: Customer Segmentation 
SELECT KMeans/DNN(prepare_feature(...)) 
FROM (
  SELECT YEAR(date) AS invoice_year,
         quantity * price AS row_price,
         or_return_quantity * price AS return_row_price
  FROM order 
  JOIN lineitem ON o_order_id = li_order_id
  JOIN order_returns ON or_order_id = o_order_id
  JOIN product ON p_product_id = li_product_id
)
[WHERE DATE_FILTER | STORE_FILTER | PRODUCT_FILTER | DEPARTMENT_FILTER];

// Template 10: Customer Satisfaction Prediction
SELECT DNN(prepare_feature(...)) 
FROM customer 
CROSS JOIN product 
JOIN lineitem ON li_product_id = p_product_id 
[WHERE CUSTOMER_INFO_FILTER | DEPARTMENT_FILTER | PRICE_FILTER];
\end{lstlisting}




\end{document}